\documentstyle[aas2pp4,astrobib,epsf]{article}
\def\lastrev{ Submitted to ApJ, 24 June 1996}

\let\oldfootsep=\footnotesep
\def\ie{{\it {\frenchspacing i.{\thinspace}e. }}}

\def\simlt{\hbox{ \rlap{\raise 0.425ex\hbox{$<$}}\lower 0.65ex\hbox{$\sim$} }}
\def\simgt{\hbox{ \rlap{\raise 0.425ex\hbox{$>$}}\lower 0.65ex\hbox{$\sim$} }}

\def\that{{\hat t}}
\def\pri{^{\, \prime }}

\def\Ntil{\widetilde{N}}
\def\flim{f_{\rm lim}}

\def\Mlim{M_{\rm lim}}
\def\VEV#1{\left\langle #1\right\rangle}

\def\etal{{\it et al.}}
\def\ie{{\it i.e.}}

\def\msun{M_\odot}

\def\umin{u_{\rm min}}

\def\Amax{A_{\rm max}}

\def\that{{\widehat t}\,} 
\def\ndf{N_{\rm dof}}
\def\Nexp{N_{\rm exp}}
\def\Nobs{N_{\rm obs}}
\def\Nshort{N_{<20}}

\def\leaderfill{\leaders\hbox to 1em{\hss-\hss}\hfill}

\def\etal{{\it et al.}}

\def\kms{{\rm\, km/s}}
\def\kpc{{\rm\, kpc}} 
\def\pc{{\rm\, pc}}

\def\ten#1{\times 10^{#1}} 
\def\msun {{ \rm \, M_\odot}}

\def\Amax{A_{\rm max}} 
\def\umin{u_{\rm min}} 
\def\tmax{t_{\rm max}} 
\def\eff {{\cal E}}

\def\ie{{\it i.e.}}
\def\lsim{\mathrel{\mathpalette\@versim<}}
\def\gsim{\mathrel{\mathpalette\@versim>}}
\def\@versim#1#2{\lower0.2ex\vbox{\baselineskip\z@skip\lineskip\z@skip
  \lineskiplimit\z@\ialign{$\m@th#1\hfil##\hfil$\crcr#2\crcr\sim\crcr}}}
\def\spose#1{\hbox to 0pt{#1\hss}}
\def\simlt{\mathrel{\spose{\lower 3pt\hbox{$\mathchar"218$}}
     \raise 2.0pt\hbox{$\mathchar"13C$}}}
\def\simgt{\mathrel{\spose{\lower 3pt\hbox{$\mathchar"218$}}
     \raise 2.0pt\hbox{$\mathchar"13E$}}}

\begin{document}

\title{The MACHO Project LMC Microlensing Results from the First Two Years
and the Nature of the Galactic Dark Halo}

\author{
	  C.~Alcock\altaffilmark{1,2}, 
	R.A.~Allsman\altaffilmark{3},
	  D.~Alves\altaffilmark{1,4},
	T.S.~Axelrod\altaffilmark{5},
	A.C.~Becker\altaffilmark{2,6},
	D.P.~Bennett\altaffilmark{1,2,4,7},
	K.H.~Cook\altaffilmark{1,2},
	K.C.~Freeman\altaffilmark{5},
	  K.~Griest\altaffilmark{2,8},
	  J.~Guern\altaffilmark{2,8},
	M.J.~Lehner\altaffilmark{2,8},
	S.L.~Marshall\altaffilmark{1,2},
	B.A.~Peterson\altaffilmark{5},
	M.R.~Pratt\altaffilmark{2,6,9},
	P.J.~Quinn\altaffilmark{10},
	A.W.~Rodgers\altaffilmark{5},
	C.W.~Stubbs\altaffilmark{2,6},
	  W.~Sutherland\altaffilmark{11},
	  D.L.~Welch\altaffilmark{12}
	}
\begin{center}
{\bf (The MACHO Collaboration) }\\
\lastrev   
\end{center}

\begin{abstract} 
\rightskip = 0.0in plus 1em

The MACHO Project is a search for dark matter in the form
of massive compact halo objects (Machos).
Photometric monitoring of millions of stars in
the Large Magellanic Cloud (LMC), Small Magellanic Cloud (SMC), 
and Galactic bulge is used to search for gravitational
microlensing events caused by these otherwise invisible objects.
Analysis of the first 2.1 years of photometry
of 8.5 million stars in the LMC reveals
8 candidate microlensing events.
This is substantially more than the number expected ($\sim 1.1$) from lensing
by known stellar populations.
The timescales ($\that$) of the events range from 34 to 145 days.
We estimate the total microlensing optical depth towards the LMC
from events with $2 < \that < 200$ days to be
$\tau_2^{200} = 2.9 ^{+1.4}_{-0.9} \ten{-7}$
based upon our 8 event sample. This exceeds the  
optical depth, $\tau_{\rm backgnd} = 0.5 \ten{-7}$, expected from known
stars, and the difference is to be compared with the optical depth  
predicted for a ``standard" halo composed entirely of Machos:
$\tau_{halo} = 4.7\ten{-7}$.
To compare with Galactic halo models, we perform likelihood analyses 
on the full 8 event sample and a 6 event subsample (which allows for
2 events to be caused by a non-halo ``background"). 
This gives a fairly model independent estimate of the halo mass in Machos
within 50 kpc of $2.0^{+1.2}_{-0.7} \ten{11} \msun$, which
is about half of the ``standard halo"
value. We also find a most probable Macho mass of $0.5^{+0.3}_{-0.2}\msun$, 
although this value is strongly model dependent. 
Additionally, the absence of short duration
events places stringent upper limits on the contribution of
low-mass Machos: objects from $10^{-4} \msun$ to $0.03 \msun$
contribute $\simlt 20\%$ of the ``standard" dark halo. 

\end{abstract}


\altaffiltext{1}{Lawrence Livermore National Laboratory, Livermore, CA 94550\\
	Email: {\tt alcock, alves, bennett, kcook, stuart@igpp.llnl.gov}}

\altaffiltext{2}{Center for Particle Astrophysics,
	University of California, Berkeley, CA 94720}

\altaffiltext{3}{Supercomputing Facility, Australian National University,
	Canberra, ACT 0200, Australia \\
	Email: {\tt robyn@macho.anu.edu.au}}

\altaffiltext{4}{Department of Physics, University of California,
	Davis, CA 95616 }

\altaffiltext{5}{Mt.~Stromlo and Siding Spring Observatories,
	Australian National University, Weston, ACT 2611, Australia  \\
	Email: {\tt tsa, kcf, peterson, alex@mso.anu.edu.au}}

\altaffiltext{6}{Departments of Astronomy and Physics,
	University of Washington, Seattle, WA 98195\\
	Email: {\tt becker, mrp, stubbs@astro.washington.edu}}

\altaffiltext{7}{Department of Physics, University of Notre Dame,
	Notre Dame, IN 46556 }

\altaffiltext{8}{Department of Physics, University of California,
	San Diego, La Jolla, CA 92093\\
	Email: {\tt kgriest, jguern, mlehner@ucsd.edu }} 

\altaffiltext{9}{Department of Physics, University of California,
	Santa Barbara, CA 93106 }

\altaffiltext{10}{European Southern Observatory, Karl-Schwarzchild Str. 2,
D-85748, Garching, Germany \\
	Email: {\tt pjq@eso.org}}

\altaffiltext{11}{Department of Physics, University of Oxford,
	Oxford OX1 3RH, U.K. 
	Email: {\tt w.sutherland@physics.ox.ac.uk}}

\altaffiltext{12}{Dept. of Physics \& Astronomy, 
   McMaster University, Hamilton, Ontario, Canada L8S 4M1. 
	Email: {\tt welch@physics.mcmaster.ca}}

\setlength{\footnotesep}{\oldfootsep}

\section{Introduction}
\label{sec-intro}

\def\yrone {A96} 

Following the suggestion of \citeN{pac86},
many groups are now engaged in searches for
dark matter in the form of massive compact halo 
objects (Machos) using gravitational microlensing, 
and many candidate microlensing events have been reported. 
Detailed reviews of microlensing are given by \citeN{roulet}
and \citeN{pac-annrev}.

The expected microlensing rate towards the Magellanic Clouds
due to  a Macho-dominated halo comfortably exceeds the
expected microlensing rate from known populations of
low-mass stars. Thus, our LMC survey
directly probes the 
Macho content of the halo. 
We have previously reported 4 microlensing candidates towards the LMC 
(\citeNP{macho-nat}, \citeyearNP{macho-iauc6095,macho-prl}, 1996a, hereafter 
\yrone); 
the EROS group has reported
2 candidates \cite{eros-nat}, but 1 of these is an eclipsing 
binary star \cite{eros-followup} and hence is suspect. 
These events have 
characteristic timescales $\that \sim 20 - 60$ days;  
searches for short-timescale events with
1 hour $\lesssim \that \lesssim$ 
10 days  have revealed no candidates to date
\cite{eros-ccd,macho-spike} and set interesting limits on low-mass Machos.  
 
The results from the two groups are consistent after accounting 
for the different sample sizes and detection efficiencies; 
a robust result is that 
substellar Machos from $10^{-6}$ to $10^{-2} \msun$ 
cannot make up the entire `standard' halo mass of 
$4.1\ten{11} \msun$ within $50\kpc$. 

However, it is very interesting that 
the number of previously detected LMC
 events appears significantly higher 
than expected from lensing by known stellar populations;  
thus, there were a range of plausible explanations 
outlined by \yrone. These included
a halo containing $\sim 20\%$ Machos, a `minimal' all-Macho halo, 
Machos in a thick disk or spheroid, a statistical fluctuation 
in the stellar lensing rate, or variable stars masquerading as 
microlensing events. 

A larger sample can help to distinguish between these alternatives. 
In this paper we present analysis of our first 2.1 years of 
observations of 8.5 million stars in the LMC; this dataset
comprises the same set of stars analyzed in 
\citeN{macho-prl} and \yrone, but with twice the timespan 
and improved selection criteria.  

The role of the unexpectedly large number of events (currently $> 100$)
seen in the direction of the Galactic bulge
\cite{ogle-tau,macho-bulge1,macho-bulge2,duo}  is indirect. 
While these results are very interesting as a verification of 
microlensing, as a  probe of Galactic
structure and the mass functions of the disk and bulge, 
and as a possible avenue for detection of planets 
\cite{mao-pac,gould-loeb,BennettRhie,tyt95},
bulge microlensing does not directly probe halo dark matter.
Our experience with bulge microlensing has taught us much, however,
and has helped us to refine our event selection criteria.

The plan of the paper is as follows: in \S~\ref{sec-obs} we outline the 
observations and photometric reductions, and in \S~\ref{sec-det} we 
describe microlensing event selection, and
the resulting candidates;  
in \S~\ref{sec-eff} we estimate our detection efficiency.
In \S~\ref{sec-dist} we show the distributions of the selected
events in the color-magnitude diagram, on the sky, and according
to impact parameter.
In \S~\ref{sec-implic} we provide various analyses of the sample, 
using both the total number of events and the individual timescales. 
We discuss our conclusions in \S~\ref{sec-discuss}. 
 
Note that many of the reduction and analysis 
procedures used here are very similar to those
in \yrone, to which we refer extensively. 
The reader is encouraged to consult that paper 
to understand the details of the experiment, but we repeat the 
main points here for clarity.  
 
\section{Observations and Photometric Reductions} 
\label{sec-obs}
The MACHO Project has had full-time use of the $1.27$-meter telescope at 
Mount Stromlo Observatory, Australia, since 1992 July;
an extended run until 1999 December has recently been approved.
Details of the telescope system are given by
\citeN{macho-tel} 
and of the camera system by \citeN{macho-stubbs} 
and \citeN{macho-marshall}. 
Briefly, corrective optics and a dichroic are used to 
give simultaneous imaging of a $42 \times 42$ arcmin$^2$ field 
in two colors, using eight $2048^2$ pixel CCDs. 
 
As of 1996 June, over 44000 exposures have been taken with the system, 
over 3 TBytes of raw image data.  About  
$60\%$ are of the LMC, the rest are of fields in the Galactic center and
SMC. 

In this paper, we consider the first 2.1 years of data from 
22 well-sampled fields,  located in the 
central  $\sim 5^o \times 3^o$ 
of the LMC; field centers are listed in Table~\ref{tab-fields}, 
and shown in Figure~\ref{fig-lmc}. 
\placefigure{fig-lmc} 

\begin{figure*} 
\epsscale{2.0}   
\vskip 7truein
\caption{An R-band image of the LMC, $8.2$ degrees on a side
(G. Bothun, private communication), 
showing the locations of the 22 MACHO fields used here. 
  \label{fig-lmc} } 
\end{figure*} 

The observations described here comprise 10827 images 
distributed over the 22 fields. These include virtually all of
our observations of these fields in the time span of 769 days from
1992 September 18 to 1994 October 26 as well as a fraction of our observations
taken between 1992 July 22 and 1992 August 23 when our system was still in
an engineering phase.
We obtained at least one observation on 556 of the 769 nights. 
The mean number of exposures per field is $10827/22 = 492$,  
with a range from 300 to 785. The sampling varies  
between fields (Table~\ref{tab-fields}), 
since the higher priority fields
were often observed twice per night with a $\sim 4$ hour spacing. 
There was a 50-day gap in observations after 1993 November 29, 
following a fire in an electronics box in the control room. 
The telescope and 
CCD cameras were not affected, so there is no systematic difference
in the data before and after the fire. 

The photometric reduction procedure was very similar to 
that described in \yrone; briefly, a good-quality image of each 
field is chosen as a template, and used to generate a list 
of stellar positions and magnitudes. 
All other images are aligned with the template using fiducial stars, 
and a PSF is measured from these. Then, the flux of all other stars
is fitted using the known positions and PSF. For each measurement
we also compute an error estimate and six quality flags; these flags
are the object type, the $\chi^2$ of the PSF fit, 
a crowding parameter, a local sky estimate, and the fraction of 
the star's flux rejected due to bad pixels and cosmic rays. 
The resulting data are reorganized into stellar lightcurves, 
and searched for variable stars and microlensing events. 

There is a minor complication to which we will refer later. 
For software-related reasons,
we used different templates for the first and
second year's reductions of 6 of our fields. Thus there
is not a one-to-one correspondence between the set of stars 
in the 2 distinct years, and the 2 years had to be 
analyzed separately. 
We chose to make this split between the pre-fire and post-fire 
data to minimize the chance of an event spanning the 2 data sets. 

\section{Event Detection}
\label{sec-det}

The data set used here comprises some 9 billion individual photometric
measurements. 
Discriminating genuine microlensing from  stellar variability and 
systematic photometry errors is not an easy task.  
The significance of the results we report in this paper
is critically dependent upon the event selection
criteria we employ.

The determination of our event selection criteria could not be made
before looking in detail at the lightcurves.
The selection
criteria should accept `true' microlensing events, and reject events
due to intrinsic stellar variability and instrumental effects.
The MACHO Project is the largest survey of astronomical variability
in history, which means that we had to learn how to perform
event selection from the data we gather; it was not possible to
develop meaningful selection criteria independent of the data.
 
The first step in the event selection process
is to use the photometry quality flags
(defined above) to reject suspect data points, and
to require a combination of good time coverage and high significance
such that systematic photometry errors do not produce false detections.
These steps leave us with lightcurves in which we can expect
to separate microlensing events from intrinsic stellar variability.
We exploit the features of
microlensing that distinguish it from
intrinsic stellar variability, and also to explicitly exclude
from our sample stars that reside in regions of the color-magnitude
diagram that are prone to variability.

We compute for each light curve a set of
temporal variability statistics, and
we have developed a set of criteria (``cuts") that we use to distinguish
microlensing from the noise.
These criteria have evolved over the course of the experiment.
The cuts used to detect the 3 LMC events reported
in \citeN{macho-prl} and \yrone\ were derived from Monte Carlo simulations
of microlensing events added to our
early data. We have refined these cuts for the analysis reported here.

In this regard, our experience with the Galactic bulge has been
especially helpful. Since our previous analysis of a smaller set of
LMC data (\citeNP{macho-prl}, \yrone),
we have analyzed a large set of data towards the
Galactic bulge \cite{macho-bulge2,macho-pratt}, which
has yielded over 80 microlensing events.  These observations 
provide a more realistic set of microlensing events than the artificial events
we have used previously to test our analysis procedures. 

Of the 43 bulge events from the 1993 bulge season
\cite{macho-bulge2}, 14 would have failed the cuts used in our previous
LMC analysis. 
A number of these events fail the cuts  described in \yrone\ because
they are not well described by the ``normal" microlensing light curve
which assumes a single point lens, constant velocities, and an unblended
source star. Using the nomenclature of \citeN{macho-bulge2}, 
events 119-A and 104-C have large deviations from the
``normal" microlensing lightcurves due to a binary lens (for 119-A) and
the orbital motion of the Earth (for 104-C). Event lightcurves can also 
deviate from ``normal" if the lensed star's image is blended with another
star which is within the same seeing disk but is not lensed.
The lightcurves of events 115-A, 128-A, 159-B, and 162-B appear to indicate
significant blending which causes them to
fail the LMC year-1 cuts on the fit $\chi^2_{ml}$, the fit $\chi^2_{ml}$ in the
peak region, and/or the event chromaticity. 
Two other 1993 bulge
events also fail the fit $\chi^2_{ml}$ 
or fit ``peak" $\chi^2_{ml}$ cuts. These are
event 119-B which appears to have a couple of photometric outliers which
barely pass our photometric quality cuts, and event 128-B which has
excess scatter in the unmagnified portion of the lightcurve for reasons
that are not yet understood.
We also find that events 101-D, 111-A, 111-B,
124-A, and 159-A fail our cuts on lightcurve coverage while event 104-B
fails our LMC year-1 crowding cut.

Many of these bulge events which fail the LMC year-1 cuts, such as events
101-D, 119-B, 128-A and 159-B 
as well as the exotic microlensing events 104-C and
119-A appear by eye to be very high quality microlensing events. In contrast,
LMC events 2 and 3 of \cite{macho-prl} which passed the 
year-1 cuts appear far less striking. 
Our data clearly indicate that these stars have brightened, but the
signal-to-noise for these events is too low to say much about the shape
of their lightcurves. Thus, we are more confident that these bulge events
are actual microlensing events than we are that LMC events 2 and 3 were
caused by microlensing. This has motivated us to modify our selection criteria
for this combined year 1 and 2 data. As described below, we have modified our
cuts so as to allow more ``high quality" events similar to the events seen
towards the bulge.

Note that as long as the experiment's event
{\it detection efficiency} is calculated properly,
and the {\it selection criteria}
are sufficiently stringent to accept only real microlensing events,
changes in the selection criteria will be accounted for in 
the efficiency calculations, and 
the details will not affect the final results, \ie\ the microlensing optical 
depth, the halo mass, and macho mass estimates.

We have made the following changes to the selection criteria.
We have loosened a cut on stellar crowding which removed stars
which are severely blended with their neighbors
under typical seeing conditions,
and we have loosened our cuts on the
stellar object type defined by our photometry code.
We loosened a cut on the average photometric error,
allowing sensitivity 
to high magnification events on very faint stars.
We have also dropped our previous cut on event chromaticity. Although
theoretical microlensing events must be achromatic,
our target stars are often blended with stars of
different colors, so observed microlensing events
can appear to have some color dependence.  (There is a generalization
of the chromaticity test that allows for blending. It is used
below to check each selected event, but we do not use
this procedure for event selection.)
 
As a result of these changes, 
we have found it necessary to increase some of our signal-to-noise
cuts to compensate for the increased background 
that the above changes allow. This reduces the chances of detecting
low signal-to-noise lensing events.

Another change in our analysis procedure was motivated by the
fact that $\sim 5$\% of the microlensing events seen toward the bulge
show significant deviations from the normal point source, point lens,
constant velocity microlensing light curve. 
These effects include binary lenses 
\cite{ogle7,macho-bennett94,macho-pratt,duo2}, an event
showing asymmetry due to the orbit of the Earth 
\cite{macho-parallax}, and an event showing the effect of the finite
source size \cite{macho-9530}. These are often quite spectacular
examples of gravitational microlensing events, but they are poorly fit
by a normal microlensing light curve. As a result, cuts on the $\chi^2_{ml}$
of the normal microlensing event fit can
preferentially remove these exotic
events resulting in a microlensing detection
efficiency that is lower than
predicted based upon a set consisting of ``normal" events. 
In the present analysis, 
we have taken some steps to reduce our dependence on $\chi^2_{ml}$
in the region of lensing magnification, and this has paid off with the
discovery of a binary microlensing event,
but our efficiency for detecting
exotic microlensing events is probably still somewhat lower than for
``normal" lensing events.

We have summarized the old and new selection criteria in 
Table~\ref{tab-cuts}; these are described in more detail below. 

\subsection{Selection Criteria}
\label{sec-cuts}

Before starting the microlensing search, measurements with questionable
PSF chi squared, crowding, missing pixel, or cosmic ray flags are flagged as
suspect measurements and removed from further consideration. 
We require that stars have an acceptable template measurement, and 
at least 7 simultaneous red-blue data pairs, to be searched for 
microlensing. 
The reddest  $\approx 0.5\%$  of 
stars, those with $V-R > 0.9$, are excluded from the microlensing search 
because they are usually long-period variables. 
The event detection then proceeds in two stages.  The first stage, 
to define a loose collection of candidate events, is very similar to 
that described in \yrone; a set of matched filters
of timescales 7, 15 and 45 days is run over each lightcurve.
If after convolution, a lightcurve shows a peak above a pre-defined
significance level in both colors,  it is defined as a `level-1 candidate',  
a full 5-parameter fit to microlensing is made, 
and many statistics describing the significance
of the deviation, goodness of fit, etc. are calculated. 
We use the standard point-source, point-lens approximation, 
in which the magnification $A$ is given by \cite{refsdal} 
\begin{eqnarray} 
  A(u) & = {u^2 + 2 \over u \sqrt{u^2 + 4} } \nonumber \\
\label{eq-amp}
    u & = b / r_E  \\
  r_E & = \sqrt{4Gm l (L-l) \over c^2 \,L } \nonumber
\end{eqnarray}
where $b$ is the separation of the lens from the undeflected line of sight,
$r_E$ is the Einstein radius (in the lens plane), 
 $l$, $L$ are the distances to the lens and source, 
and $m$ is the lens mass. 
Assuming also uniform motion, 
 the fit to microlensing takes the form 
\begin{eqnarray} 
  f_R(t) & = A(t) f_{0R}, \qquad f_B(t) = A(t) f_{0B} \nonumber \\ 
\label{eq-mlfit}
  A(t) & = A(u(t)) \\
  u(t) & = \left[ \umin^2 + 
    \left( { 2 (t - \tmax) \over \that } \right)^2 \right]^{0.5} \nonumber 
\end{eqnarray}
 where the 5 free parameters are 
 the baseline flux in red and blue passbands  $f_{0R}, f_{0B}$, 
 and the 3 parameters of the microlensing event: 
 the minimum impact parameter in units of the Einstein radius $\umin$ , 
 the Einstein diameter crossing time $\that \equiv 2 r_E / v_{\perp}$,
 and  the time of maximum magnification $\tmax$. 
 Later, we will often quote the more observable
 fit maximum magnification $\Amax \equiv A(\umin)$ instead of $\umin$. 

Lightcurves passing loose cuts on these statistics 
 are defined as 
 `level-1.5' candidates, and are output as individual files along
 with their associated statistics. 
 In the present analysis, there were approximately 390,000 level-1 candidates, 
 of which about 29,000 passed the level-1.5 criteria. 
 More stringent cuts are then applied
 to select final `level-2' candidates.

As explained above, the criteria for `final' microlensing candidates have been 
modified from those used in \yrone. An important 
parameter that is used for a number
of the cuts is $\Delta \chi^2 \equiv \chi^2_{\rm const} - \chi^2_{ml}$ 
where $\chi^2_{\rm const}$ and $\chi^2_{ml}$ are the $\chi^2$ values for the
constant flux and (unblended)
microlensing fits respectively; $\Delta \chi^2$ is
 the effective `significance' of the event summed over all data points. 
$\chi^2_{\rm peak}$ refers
to the $\chi^2$ of the microlensing fit in the ``peak" region where
$A_{\rm fit} > 1.1$. $\ndf$ refers to the number of degrees of freedom
in a particular fit.
 We use the following criteria to select candidate microlensing events: 

\begin{description}

\item[1)] The fitted time of peak magnification 
 $\tmax$ must be within the time span of the observations,
 and the event duration $\that < 300\,$days.

\item[2)] We require that the star should show a roughly constant 
 baseline: 
 there must be at least 40 `baseline' points 
 outside the time interval $\tmax \pm 2\that$, 
 and the reduced $\chi^2$ of the microlensing fit outside this interval 
 must be $\chi^2_{ml-out}/\ndf < 4$.

\item[3)] We require that $A_{\rm max} > 2\times$ the mean 
  estimated error of the data points. 

\item[4)] We require at least 6 data points $> 1\sigma$ 
 above median brightness in the peak region $\tmax \pm 0.5\that$. 
 This excludes candidate events which may be caused by 
 one or two discrepant observations, most commonly un-flagged cosmic rays, 
 satellite tracks, poor telescope tracking etc.  

\item[5)] We exclude stars brighter than  $V < 17.5$ which contain 
 a class of bright blue variables known as ``bumpers" \cite{macho-bump}.

\item[6)] We exclude stars in a $10' \times 10'$ 
 region surrounding SN 1987A in order to
  avoid spurious microlensing triggers  due to the supernova light echo.

\item[7)] We remove events with low signal to noise or a poor peak fit 
by requiring  $\Delta\chi^2/(\chi^2_{\rm peak} / \ndf) > 200$.

\item[8)] We remove stars that may have crowding related spurious photometry;
 those stars for which over 5\% of the measurements have been rejected due
 to crowding. 

\item[9)] We require a crowding dependent signal/noise criteria:
 crowding measure  $< 70 \log_{10}[\Delta\chi^2/(\chi^2_{ml} / \ndf)]-45$.
 This cut requires extra signal-to-noise for stars with very close neighbors 
 and removes a class of spurious triggers caused by PSF anomalies due to
 transient problems with the telescope optics.

\item[10)] Our main signal-to-noise cut is 
 $\Delta\chi^2/(\chi^2_{ml} / \ndf) > 500$.

\item[11)] We require a fit $A_{\rm max} > 1.75$.

\end{description}

\begin{figure*}
\epsscale{2.0}
\plotone{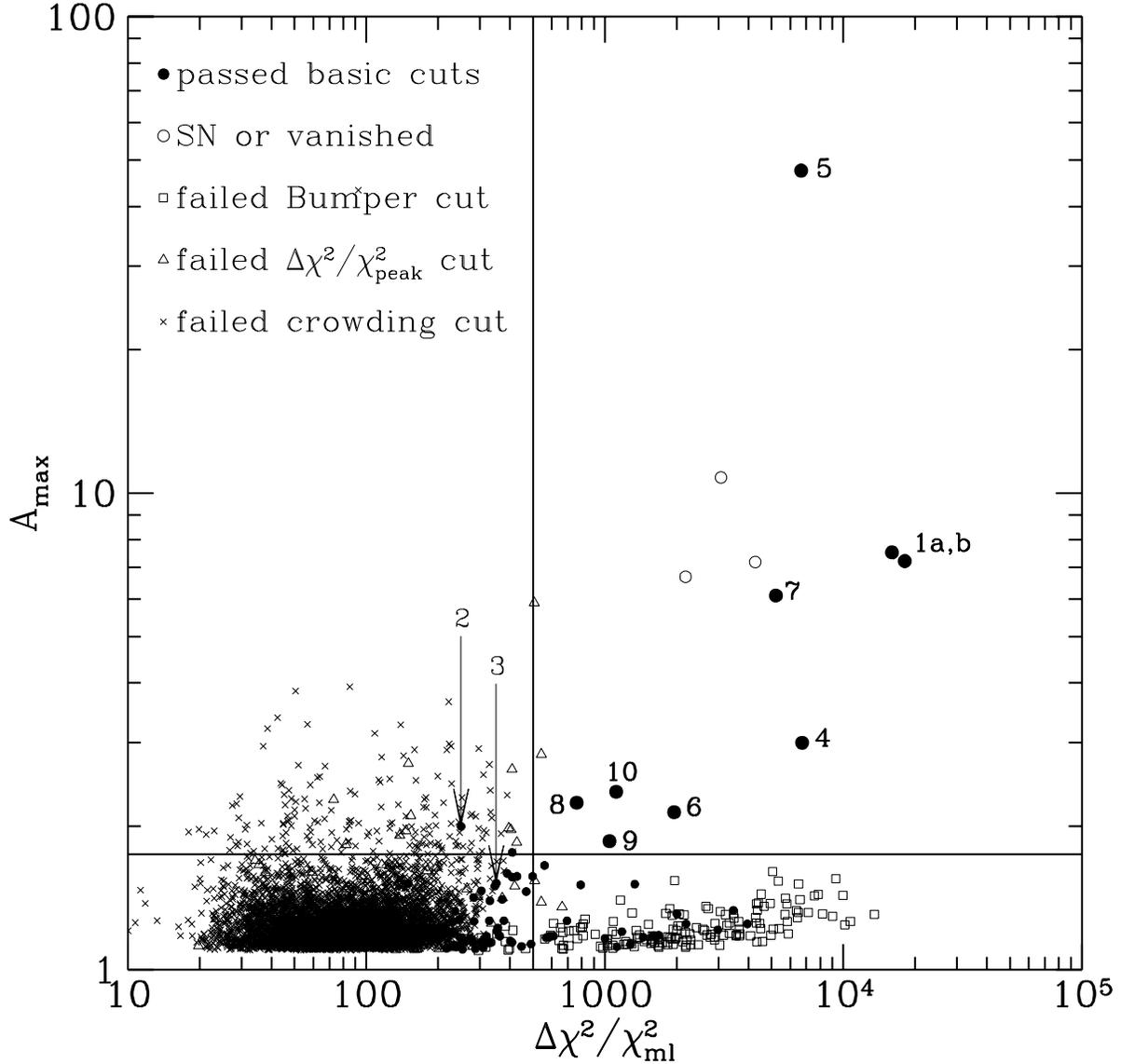}
\caption{The final cuts for selection of microlensing candidates. 
 The x-axis is $\Delta\chi^2/\chi^2_{ml}$, where
 $\Delta\chi^2 \equiv \chi^2_{const} - \chi^2_{ml}$ is the improvement
 in $\chi^2$ between a constant brightness fit and a microlensing fit.
 The y-axis is the fitted maximum magnification. 
 The symbols are explained in \S~3.1. 
 The solid lines show the final cuts (10) and (11); the circles
 in the upper right are the 12 events (10 stars) discussed in 
 \S~\protect\ref{sec-events}.  
 Events~2 and 3 from \yrone\ are also indicated. 
 \label{fig-cuts} } 
\end{figure*}

These cuts are summarized in Figure~\ref{fig-cuts}, which illustrates cuts 
 (10) and (11) for all  events that pass cuts (1)-(4) and (6). 
\placefigure{fig-cuts} 

Events that pass all cuts 
except for (10) and (11) are indicated with circles, while the
open symbols and crosses indicate events which failed 1 or more of
cuts (5), (7)-(9). Open squares indicate events which fail cut (5), and 
open triangles indicate events which fail cut (7). The crosses indicate
events which fail cuts (8) or (9).
Note that while most of these events were discovered
prior to the selection of the final set of cuts, the event closest to
the cut boundaries was not discovered until this set of cuts was run.
This is the binary lens event with a single lens fit $A_{\rm max} = 1.86$.
The lightcurves passing all of these selection criteria are further
 investigated, as outlined below. The effect of the variation of these
cuts on our results are discussed in Section~\ref{sec-tau_cut} below.

\subsection{Microlensing Candidates} 
\label{sec-events}

\begin{figure*}
\epsscale{2.0}
\plotone{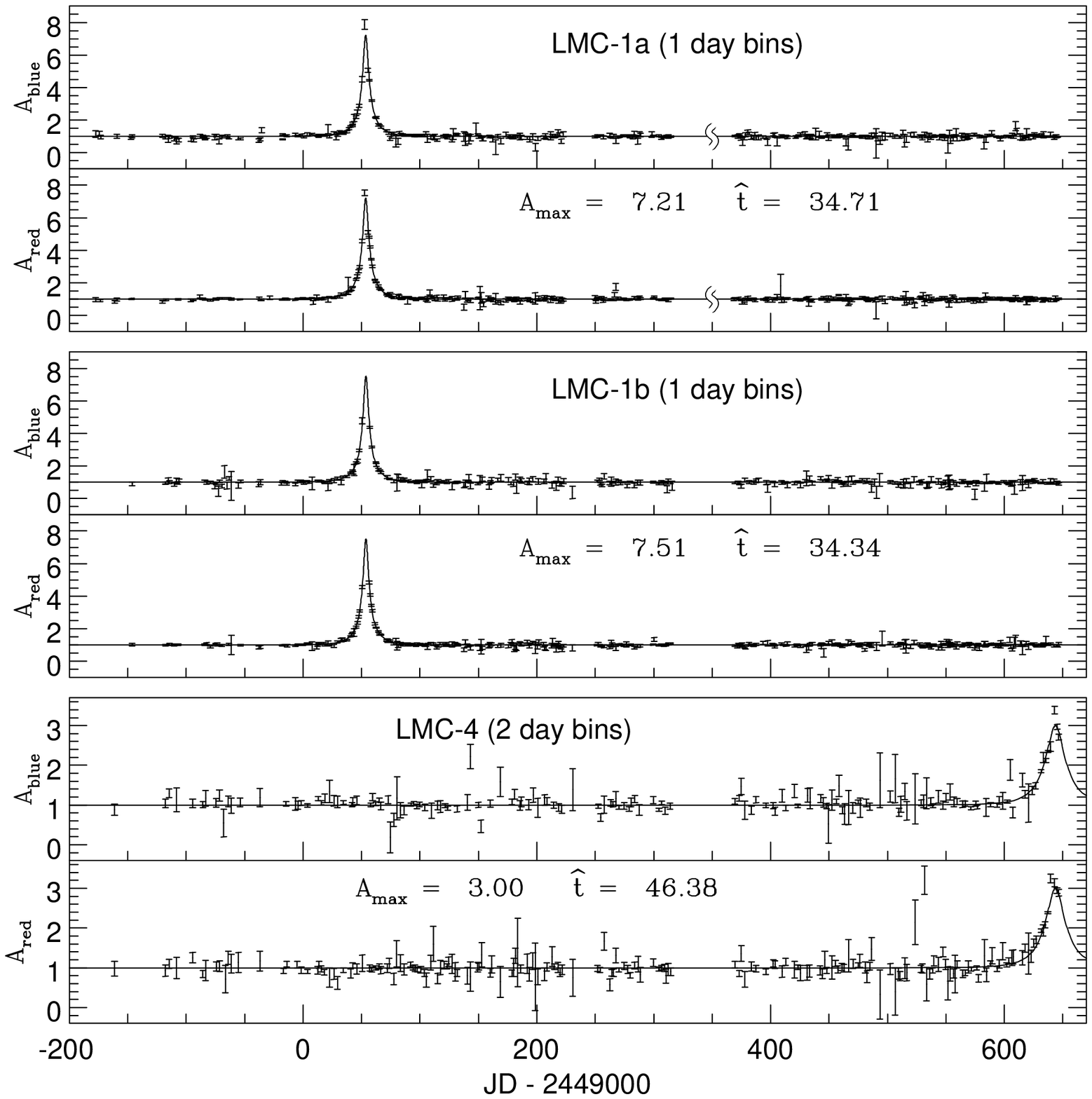}
\caption{The lightcurves for the 12 candidates (10 stars) in 
 \S~\protect\ref{sec-events}.
 For each object, the upper and lower panels show blue and red passbands. 
 Flux is in linear units with $1\sigma$ estimated errors, 
 normalized to the fitted unlensed  brightness. 
 For clarity, the points shown are averages in time bins roughly matched
 to the event timescales, as indicated on each panel. 
 For events 1a, 6 and 8, the wavy lines indicate that different
 templates were used before and after day 330; 
 thus a separate baseline normalization has been used for each portion; 
 see \S~\protect\ref{sec-obs} . 
 \label{fig-events} }
\end{figure*}

Twelve lightcurves passed the 
cuts discussed above, indicated by the
open and closed circles in the upper right hand region of 
Figure~\ref{fig-cuts}. 
The lightcurves are shown in Figure~\ref{fig-events}, 
their fit parameters are listed in Table~\ref{tab-events}, 
and finding charts for each star are shown in Figure~\ref{fig-fchart}. 
Parameters for fits including the possibility of blending with an unlensed star 
in the same seeing disk as the lensed star are given in Table~\ref{tab-blend}.
Four of these lightcurves (1a, 1b, 12a and 12b)
actually correspond to two stars which occur in field overlap
regions; the two lightcurves for each star 
are based on independent data and reductions. 

\placefigure{fig-events}
\placefigure{fig-fchart}

\begin{figure*}
\epsscale{2.0}
\plotone{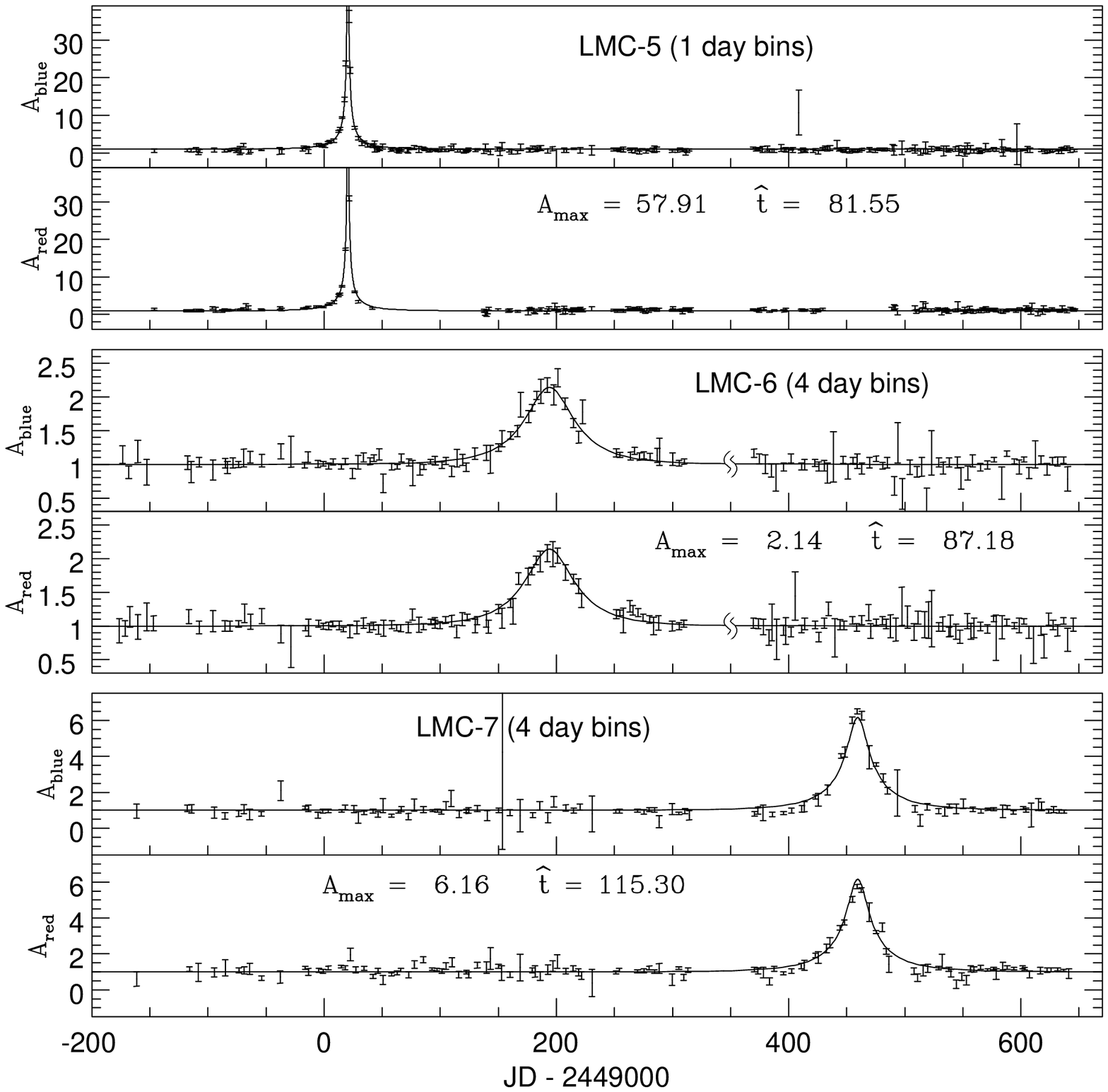}

Figure~\protect\ref{fig-events} (continued)
\end{figure*}

\begin{figure*}
\epsscale{2.0}
\plotone{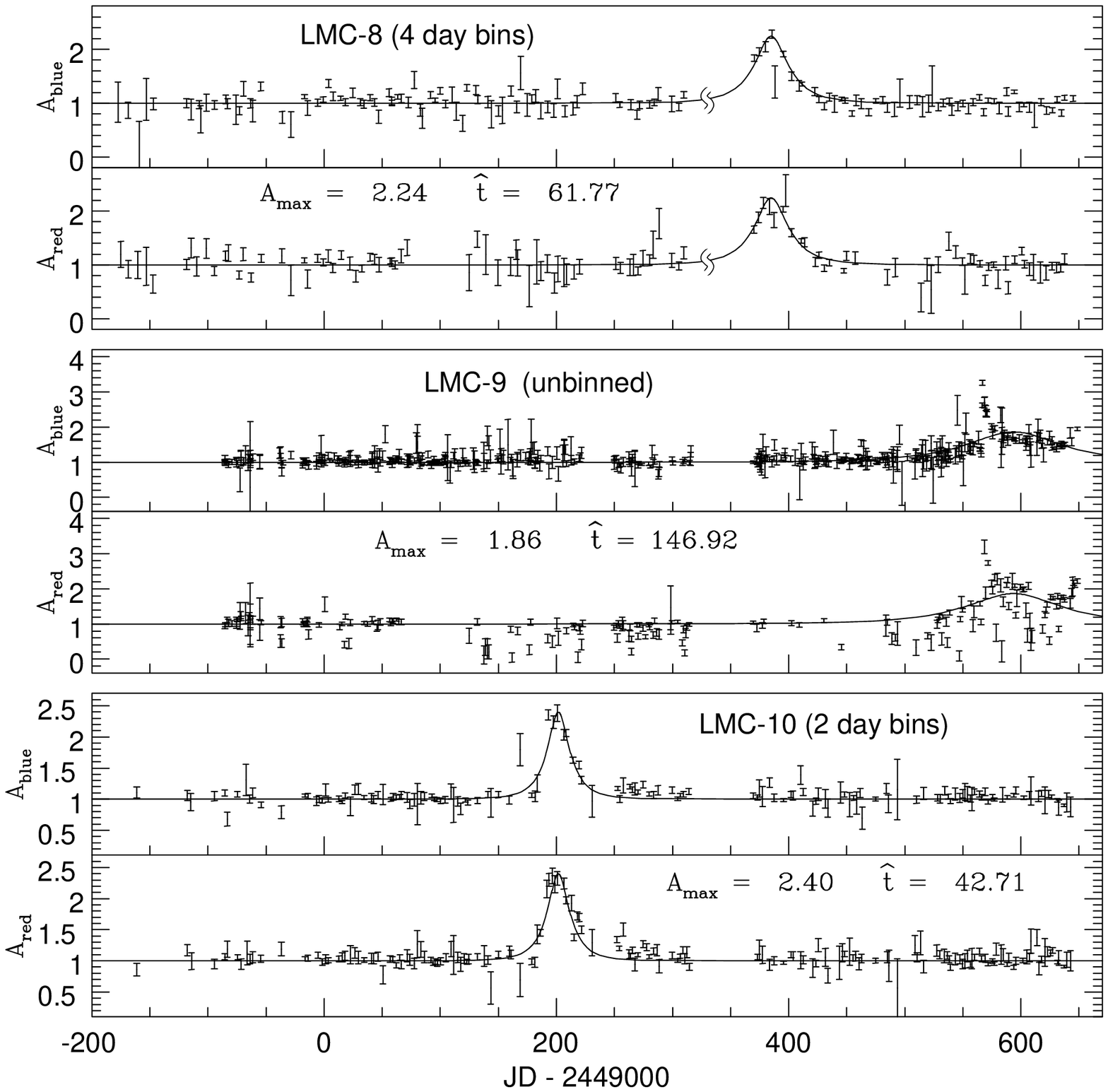}

Figure~\protect\ref{fig-events} (continued)
\end{figure*}

\begin{figure*}
\epsscale{2.0}
\plotone{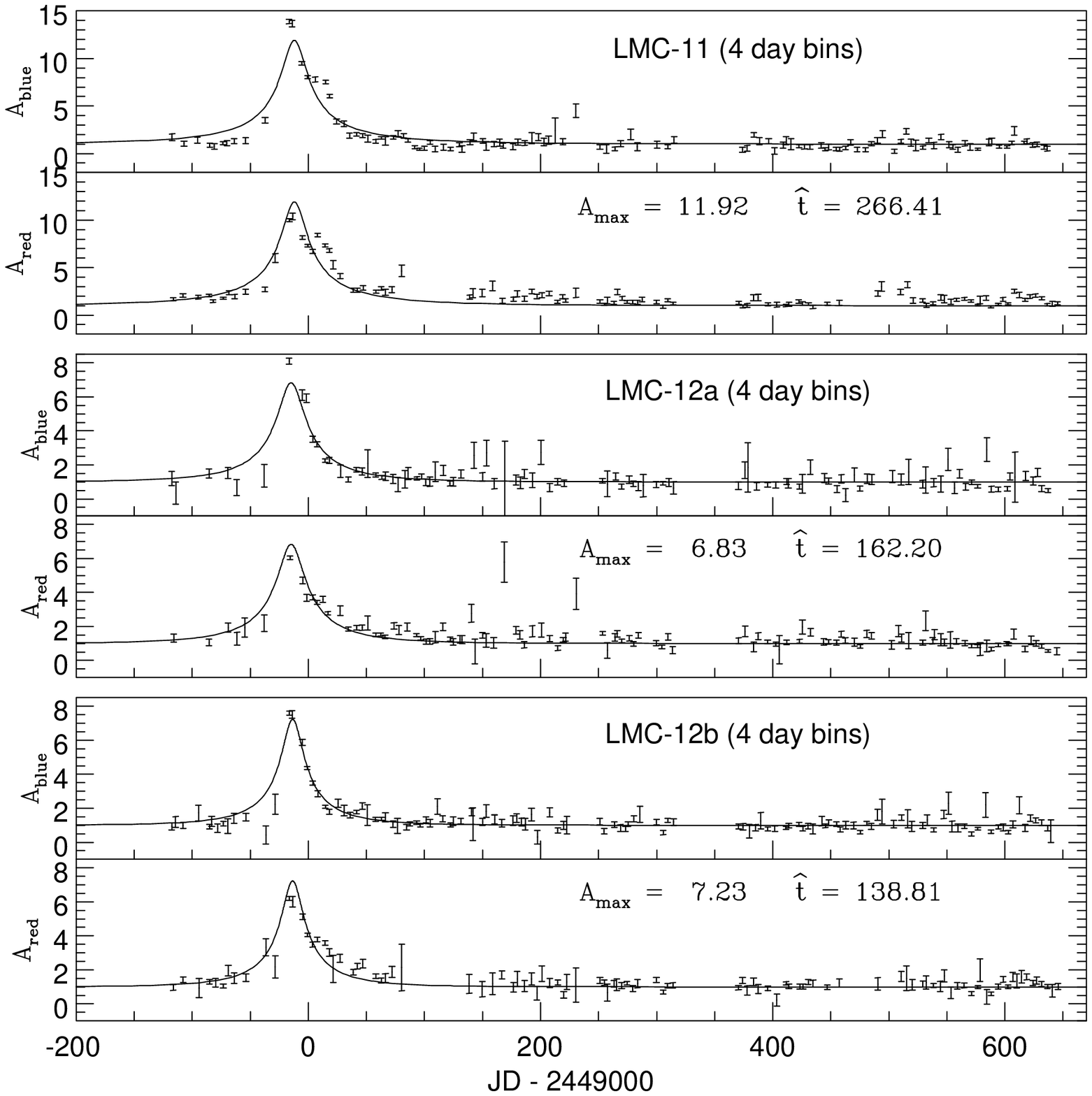}

Figure~\protect\ref{fig-events} (continued)
\end{figure*}

\begin{figure*}
\vskip 7truein 
\caption{A mosaic of 2 images centered on each candidate event; the labelled
 image shows the event near maximum observed brightness, and 
 the un-labelled image shows the event at normal  brightness. 
 These are R-band CCD images, with a scale of $0.63''$/pixel. 
 Each box is $40$ arcsec square; North is up and East is left. 
 \label{fig-fchart} }
\end{figure*}

We are very confident that these events are of astrophysical origin
and not due to systematic photometry errors. 
First, they show very significant changes in brightness 
over each of  many individual CCD images over many weeks. 
Due to pointing variations, 
a given star moves by many pixels between exposures, 
and may move to different chips when the telescope moves across the
mounting pier; so the events cannot be due to uncatalogued CCD defects. 
Two of these events are independently detected in  field overlaps, 
event~1 has been confirmed in the EROS plate data \cite{eros-nat}, 
and event~4 was confirmed by E.~Giraud at ESO \cite{giraud}. 

Once again the experience with the Galactic bulge is helpful.
A number of bulge events which have been observed from 
multiple sites. We re-discovered the events 
OGLE\#1 and OGLE\#7 in our data \cite{macho-bulge2}. 
OGLE rediscovered our Alert 95-11, 
Most of our $\sim 40$ bulge ``alert" events were
observed at other sites \cite{macho-alert1,macho-pratt,ogle-macho}; 
so far, none has turned out to be due to observational error, even
though the criteria for an alert
event are less stringent than those 
used for this sample. These multi-site detections convince
us that the events we describe are astrophysical in origin.

We now give a brief discussion of the individual events, 
with follow-up observations and a subjective quality 
classification. 

Event~1 was our first discovery \cite{macho-nat}, and 
has remained constant as expected in the following year. 
The star's spectrum is that of a normal clump giant 
at the radial velocity of the LMC \cite{dellavalle}. 
In view of the high signal to noise, achromaticity, 
and good fit to microlensing, 
we classify this as an excellent microlensing candidate. 

The low signal-to-noise candidates 2 and 3 from
\citeN{macho-prl} and \yrone\
{\it do not pass} the final cuts used for this data set;
thus, they are not included in  Table~\ref{tab-events},
but we have numbered the present set of 10 candidates
$1, 4\ldots12$ to avoid possible ambiguity.
As shown in  Figure~\ref{fig-cuts}, event~2 fails cut (10) and
event~3 fails cuts (10) and (11); this is due to the changes that
have been made in our cuts as noted above.
Although the star involved in event~2
appeared constant during the second year data,
the EROS group has informed us that it may have brightened in  1990,
and it has shown indication of further brightening episodes in our data
in 1995 January \& December; thus it is probably a variable star.
The star involved in event~3 has not shown further variation, and
may have been a microlensing event,
but it does not pass our revised selection criteria.

Event~4 was the first LMC microlensing candidate detected in 
progress; it was detected by our real-time Alert system 
(\citeNP{macho-pratt}) on 1994 October 8, and announced in 
IAU Circular 6095. This accounts for the increased frequency of 
observations during the event. We also obtained photometry
from CTIO, and several observers obtained additional 
photometry and spectra of the star 
near maximum \cite{giraud}. 
We classify this as an excellent microlensing candidate. 

Event~5 is a very high magnification event ($\Amax \approx 60$) 
in a faint star. Note that this event occurred during
the period examined in \citeN{macho-prl} and \yrone\, but
did {\it not} pass the cuts then employed (too crowded
and detected color change). 
Some deviations from the microlens fit are seen in the wings, and 
the event exhibits clear color variation (see
Figure~\ref{fig-cmd}). Neither phenomenon is
unexpected in such a faint star: there are possible systematic
photometry errors due to crowding when the star is near 
minimum, and the star is very probably blended with
another star of different color. This is reflected the substantial
improvement in the 
reduced $\chi^2$ from 1.680 in the fit without blending
to 0.965 in the fit including blending.
The symmetry of the lightcurve and 
the good fit around the peak strongly suggest that this is 
a microlensing event. 
We classify this as an excellent microlensing candidate. 

Event~6 is a moderate magnification candidate.
We classify this as a good microlensing candidate.

Event~7 is a moderately high amplitude microlensing event located well
away from the LMC bar. The light curve is quite symmetric, but some
small systematic deviations from the fit can be seen in the wings of
the light curve. 
These deviations are probably due to a
known systematic error of our photometry code:  
since stars are fitted in order of descending template brightness,
when the seeing is poor the flux from a faint star may be `stolen'
by a brighter neighboring star\rlap.\footnote{We are currently working
in collaboration with P.~Stetson to develop an improved photometry code
to use on detected events, which should avoid this systematic error.} 
Once the photometry code has detected a $> 7\sigma$ deviation brightness
variation of a given star, the star and all its close neighbors undergo another
iteration of fitting. This reduces the effect of this `stolen flux' problem
when the magnification is large. This star also exhibits some color
variation that can be explained by blending.
We classify this event as a good microlensing candidate. 

Event~8 is a moderate magnification microlensing 
candidate located in the LMC bar. 
It occurred just after the gap due to the fire, 
but we still have a reasonable
portion of the light curve's rise.
We classify this as a good microlensing candidate. 

\begin{figure}
\epsscale{0.9}   
\plotone{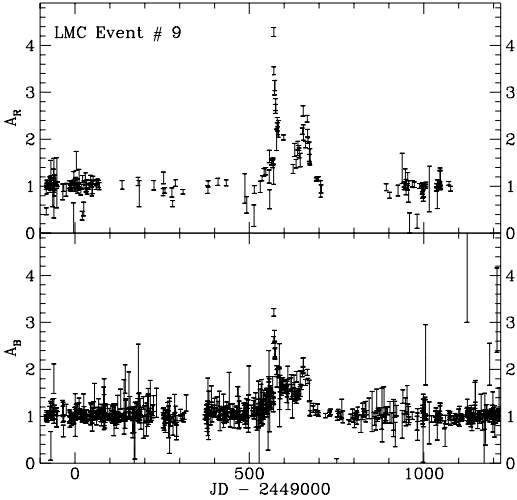}
\caption{ The full lightcurve of event~9, including data from additional 
 observations taken after 1994 October. Many of the red data points
 corrupted by a CCD trap have been removed as described in the text.
   \label{fig-bin_long} } 
\end{figure}

\begin{figure}
\plotone{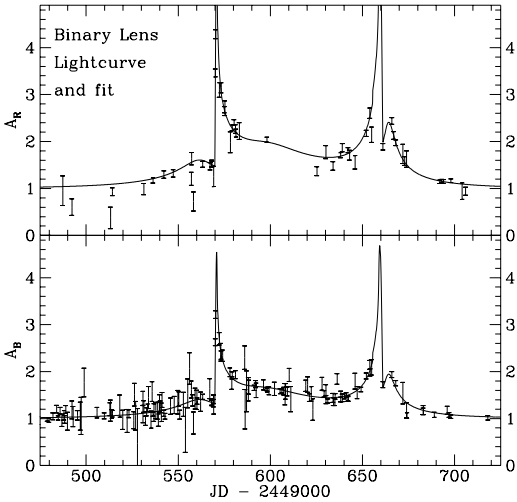}
\caption{ Lightcurve of event~9 in the peak region, 
 with the best-fit binary lens lightcurve. 
 Red data points corrupted by a CCD trap have been removed. 
   \label{fig-bin} } 
\end{figure}

Event~9 passes our objective data cuts, but it appears to be quite 
different from a normal microlensing event. 
The red baseline data contains a great deal of scatter, 
but we have determined that this is due to a previously
uncatalogued CCD defect: a low level trap that causes images to be smeared
across the column with the trap. This defect is responsible
for the scattering of low measurements in the red. In an attempt to
remove the contaminated observations, we have eliminated all the red
observations in which this defect comes within 6 pixels or 1.5 times the
image FWHM of the centroid of this star. This procedure should remove most,
but probably not all of the photometric contamination due to this
CCD defect. The corrected lightcurve of this star is shown in 
Figure~\ref{fig-bin_long}. This lightcurve includes data through early 1996,
and it was used to obtain a good fit to a binary lens light
curve \cite{macho-bennett96,macho-lmcbinary}, shown in Figure~\ref{fig-bin}. 
This fit has a $\chi^2$ of 1.76 per degree of freedom
which is quite typical of other microlensing events. 
It also requires that a significant amount of 
the total light comes from an
unlensed star in order to obtain a good fit. 
This is also the case for many of the other
binary lensing events which have been observed 
\cite{ogle7,duo2,macho-bennett94,macho-pratt}.

\placefigure{fig-bin_long}
\placefigure{fig-bin}

The multiple peaks in the lightcurve with the `U-shaped' structure 
are quite typical of binary microlensing events 
\cite{mao-pac}, but would be 
unexpected for a variable star. In particular, the gradual 
rise before the first sharp peak would be very unusual for an eruptive
variable. 
We therefore classify this as an excellent microlensing candidate. 

The two measurements on the rising portion of the first caustic
crossing (the first sharp feature in the event)
are  potentially very useful.
In principle, this will
give us an estimate of the angular velocity of the lens 
and thus a constraint on the location of 
the lens along the line of sight. Our 
analysis of this event suggests that
the lens is located in the LMC (but note that there are
two data points on the caustic crossing).
It has been estimated \cite{macho-bennett96,macho-lmcbinary} that there
is an a priori probability of about 10\% that the source star is also
a binary with parameters that would affect the apparent caustic crossing
time. The observed caustic crossing time could reflect the time it
takes the caustic to cross the orbit of a binary rather than the stellar
disk of a single star, so the possibility that this lens system is in the 
Milky Way halo cannot be ruled out.

Event 10 passes all of our cuts.  The asymmetry seen in the light
curve, a rapid rise with a more gradual fall, is typical of an eruptive
variable star. On the other hand, asymmetric light curves can also arise from 
binary lensing events and from deviations from the constant velocity 
assumption. Although event 10 is not close to any of our cut boundaries,
we consider it to be our weakest candidate because of the lack of a direct 
explanation of its light curve asymmetry. If this star is indeed a variable,
our experience suggests that it will undergo another outburst in
the future.  The inclusion or exclusion of this event
has little influence on our results. We classify it as a marginal
microlensing candidate.

Lightcurves 11 and 12a/b are indicated with open 
circles in Figure~\ref{fig-cuts} because while
they were present in our reference template images,
they are now no longer detectable in images that go somewhat deeper. 
Thus, the unbrightened images of these objects would not have been
bright enough for these stars to be detected had we selected
template images which were taken later.

Star~11 is superimposed on a background galaxy 
(see Figure~\ref{fig-fchart})
which is at a redshift of 0.021
(spectrum courtesy of J.P.~Beaulieu of the EROS collaboration).
As noted above, it was invisible on CTIO images to $V > 22$ in 
1995 December. 
Since obvious galaxies cover a very small fraction of our images 
($< 0.01\%$), this is very unlikely to be a chance superposition. 
The event was almost certainly a supernova in the background galaxy, 
probably of type II. 
There is some indication of a plateau or secondary peak about 
$25$ days after maximum. 
For $H_0 = 75 \, {\rm km/sec\, Mpc^{-1}}$, 
it had a peak brightness of about $M_V \approx -17.5$.

It is possible that event 12
is actually a microlensing event of a
star that is normally below our detection
threshold, but since we have only detected this star because it was
brighter than normal in our template frame, 
it is not proper to include it as a microlensing candidate. 
Our detection efficiency calculations do not allow for lensing
of stars which are too faint to have been detected in the template image
unless they contribute light to a 
blended composite object that is above the
threshold.

Although events~11 and~12 appear to have
significantly positive baseline 
fluxes in Figure~\ref{fig-events}, this is due to 
crowding-related systematic errors, and the presence of the 
underlying galaxy for event~11. 
If a star was present in our template image, our photometry code
will estimate its flux in all other images 
even if it is not significantly detected in them. 
The points shown in Figure~\ref{fig-events} are averages in 4-day bins; 
most of the individual data points are within $2\sigma$ of zero
flux, but the averaging tends to enhance
the apparent significance of the 
systematic errors.  

In summary, we classify events~1, 4, 5, and 9
as `excellent' microlensing
candidates, 6, 7 and~8 as `good' candidates, and~10 as a
`marginal' candidate. Events 11 and~12 must be rejected as explained
above,  
and events 2 \& 3 from \yrone\ do not pass the current cuts. 

\subsection{Comparison with Previous Cuts}
\label{sec-previous} 

It is worth noting that of the above candidates, 
numbers 1, 5, 6, 10, 11, 12 occurred in the \yrone\ data set. 
but only event 1 passed the old cuts; we discuss the others below. 
Event~5 failed cuts on crowding (1) and chromaticity (4) in the year-1
analysis.  As noted above, we expect events in faint stars to be somewhat 
chromatic due to blending, which is why the achromaticity
cut has been relaxed. 
Event~6 failed cuts on a SoDOPHOT object type for individual measurements
in the year-1 analysis. This cut was made inadvertently in the
year-1 analysis and was corrected after it was noticed in the bulge
alert events 95-5 and 95-7\rlap.\footnote{
Information on MACHO microlensing alerts can be found at
URL: {\tt http://darkstar.astro.washington.edu/} on the WWW.}
(Note that this type cut {\it was} taken into account 
properly in the year-1 LMC efficiency calculations.)
Event~10 failed year-1 cut (3) on $\chi^2_{\rm peak}$ which has been 
relaxed. Event~11 also failed the year-1 timescale cut (6) as well as the
$\chi^2_{\rm peak}$ cut, and event~12 failed the year-1 achromaticity
cut (4).

Events 10 and 12 were noted by eye as `near-miss' candidates
in our previous analysis, but the star and photometry 
quality cuts were applied at an early stage and so events
5 and 6 were not discovered in the year-1 analysis. 
Since the lightcurves of events 10, 11 and 12 differ substantially
from normal single lens microlensing, it is
not surprising that these failed the old cuts; but events
1, 5 and 6 are strong microlensing candidates, and it appears 
disconcerting that only 1 of these 3 events passed the old cuts. 
Based on our efficiency estimates below,  
we would expect the \yrone\ cuts to reject around 15\% of single-lens
events which pass the new cuts. 
The binomial probability of $\le 1$
`success' out of 3 `trials' given an 
85\% probability is $\approx 6\%$, so this appears
unlucky but not unusually so given the small-number statistics. 

\subsection{How Many Events?}
\label{sec-nevents}

The 12 lightcurves that pass our revised event selection criteria 
represent 10 objects, as two of them (events 1 and 12) occur in 
field overlap regions and were independently detected.  The supernova
(event 11) is not microlensing.  Event 12 was detected only
because of its being magnified above our object detection 
threshold in the template frame. As our detection efficiency
determination does not include this effect, event 12 will be 
excluded from further consideration. 

This leaves a set of 8 apparent events (1 and 4-10), one 
of which appears to be a binary lens (9),  that we will use to 
estimate the total optical depth towards the LMC.  As shown
below, the result we obtain exceeds the optical depth expected 
from known Galactic and LMC populations.  It is therefore useful to 
define a subset of the events that excludes lensing by known objects, 
as their distribution of timescales contains information about
the mass, velocity and spatial distributions of the new lensing
population. The difficulty is in choosing which events to 
include as members of the subsample.  Fortunately all the 
events have, in a broad sense, comparable values of $\that$
so the number of events in the subsample is more important 
than the exact choice of which ones to include. 

Given the anticipation of just over one event in the sample from 
known Galactic and LMC populations, we have elected to construct
a somewhat conservative subsample that contains 6 of the 8 events.  
The binary lensing event 
(9) was excluded for two reasons: 1) our analysis shows 
that it may come about from lensing by a binary within the LMC, and 2)
it is the longest of the events, and therefore is the most conservative
one to exclude.  We have also chosen to eliminate event 10 from 
the subsample.  This preserves the average value of $\that$ for 
the two samples.  We also think on subjective grounds that it 
is the weakest of the 8 events.  The 6 event subsample
containing events 1 and 4-8 is then a reasonable estimate of
the events from unknown lensing populations.

\subsection{Blending Effects}
\label{sec-blend} 

Many of our detected `objects' actually consist of double or
multiple stars within the seeing disk. 
This is unavoidable at the high stellar density in our fields.
Since the angular Einstein radius is much smaller than the 
seeing disk, 
only one star in a blend will typically be lensed; 
this distorts the observed lightcurve from the shape given in 
eq.~\ref{eq-mlfit}. 
We may model this by adding two parameters to represent the
flux of the unmagnified component in each color.
Specifically, 
\begin{eqnarray} 
\label{eq-blend}
  f_R(t) & = & f_{uR} + A(t) f_{lR}, \\
  f_B(t) & = & f_{uB} + A(t) f_{lB},  \nonumber 
\end{eqnarray}
where $A(t)$ is as in eq.~\ref{eq-mlfit}, $f_{uB}, f_{uR}$ are the flux of the 
unlensed component in each passband, and $f_{lB}, f_{lR}$ are the baseline 
flux of the lensed star in each passband. We use instrumental flux units
which roughly
correspond to 100 detected photo-electrons in a 300 second exposure.

The results of these fits are shown in Table~\ref{tab-blend};
we see that for
events~1, 6, 8, 10 the best fit is quite close to the unblended case, 
while for events~4,5 and 9 the best fit contains substantial blending. 
This is important because blending causes a significant
change in the parameters; 
if an event exhibits significant blending, the
single-lens fit will systematically underestimate $\Amax$ and
$\that$  relative to the true values. Clearly, we can correct for this effect
by using the blend fit values for $\that$ when we calculate the microlensing
optical depth or the most likely lens mass. These $\that$ values are the
most accurate estimate of the true $\that$ value for each event, but we
are concerned that small systematic errors in our photometry might cause us
to systematically overestimate $\that$.
Therefore, we have chosen a less accurate, but unbiased method for
estimating $\that$. We use the $\that$ values from the unblended fits
and then apply a statistical correction 
to each of our observed $\that$'s. This average correction has been
determined from our Monte Carlo simulations, and the
resulting values are shown in Table~\ref{tab-blthat}. This technique
works for all events except for the binary event which was not represented
in our Monte Carlo calculations. The blended, binary lens fit $\that$
value is used for this event.
 
We previously imposed the condition that there be no color
variation in a microlensing event (Alcock \etal\ 1993, 1994,
1995a; \yrone). A convenient graphical illustration of
the color test is a plot of $f_R(t) / f_B(t)$ {\it versus}
time. Event 1 shows no variation of this flux
ratio with time (Alcock \etal\ 1993).
Event 5, however, would show clear color variation.
One virtue of this graphical
presentation is that the test is independent of the
magnification $A(t)$.

\begin{figure*}
\epsscale{1.8}
\plotone{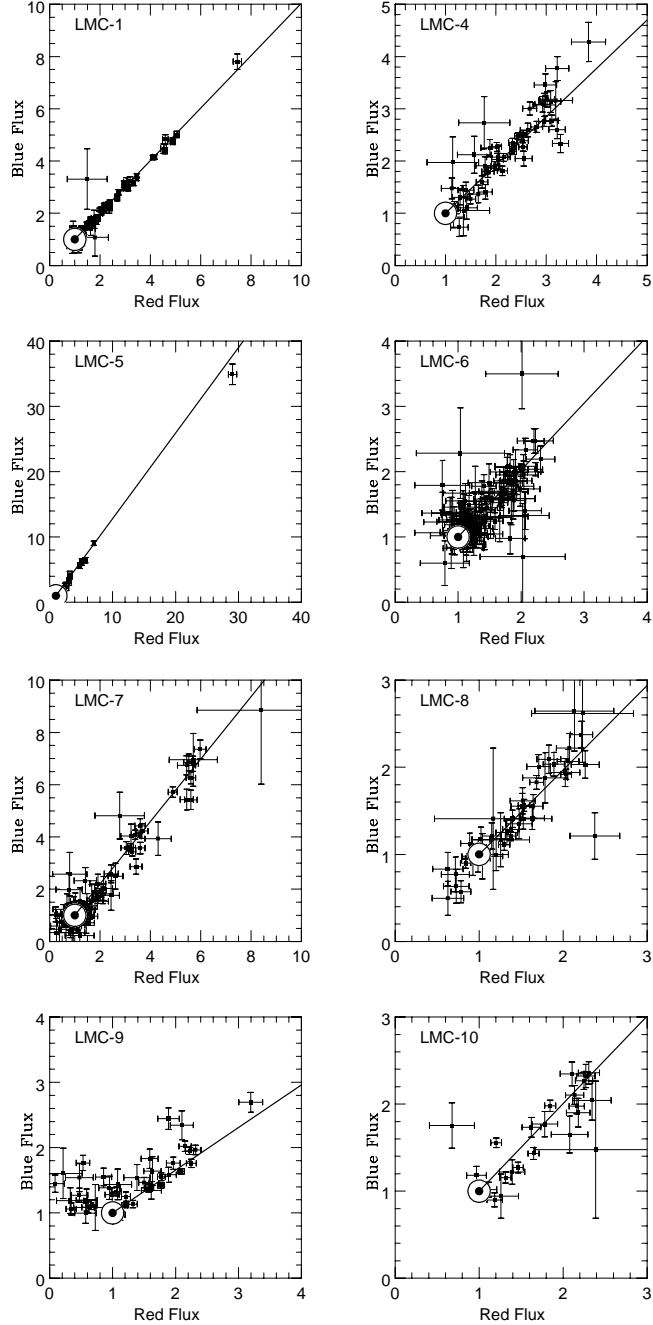}
\caption{ The blue flux is plotted against the red flux for all the 
 simultaneous measurements within 1 $\that$ for all of the events passing
 our microlensing cuts. The solid lines indicate the flux ratios
 predicted by the blended microlensing fits listed in 
  Table~\protect\ref{tab-blend}.
   \label{fig-color} }
\end{figure*}

Eq.~\ref{eq-blend} suggest an obvious generalization. A
plot of $f_B(t)$ {\it versus} $f_R(t)$ should be a straight
line. This should be true for any functional form of $A(t)$,
and can be applied equally to single, binary, or more 
general microlensing events as long as the multi-color measurements
are simultaneous. Figure~\ref{fig-color} shows
these plots for our 8 candidate events. There is no
evidence for systematic deviation from straight line
behavior in any of these plots. The straight lines
plotted in Figure~\ref{fig-color} are derived from the
best fit of eq.~\ref{eq-blend} to each event.

\section{ Detection Efficiency } 
\label{sec-eff}
 
In order to draw quantitative conclusions, we clearly need
reasonably accurate knowledge of our detection efficiency. 
The detection probability for individual events 
depends on many factors, e.g. the 3 event parameters $\Amax$, $\that$, 
$\tmax$, and the unlensed stellar magnitude, as well as our
observing strategy and weather conditions.  However, we can 
average over the known distributions in $\Amax$ and 
$\tmax$ and stellar magnitude, using the known time-sampling
and weather conditions, to derive our efficiency as a function 
only of event timescale $\eff(\that)$. 

We have computed our detection efficiency using an essentially
identical method to that outlined in \yrone, simply generating
simulated microlensing events with $\that$ logarithmically distributed
in the range 0.3--1000 days over the wider time interval, ${\rm
JD}-2,449,000 = -176.5$ to 663.5, and adding these simulated events
into the extended timespan of observations. The Monte-Carlo procedure
takes into account the actual spacing and error bars of the
observations, so any variations in sampling frequency, weather, seeing
etc.~between the first and second year data are automatically
accounted for.

There are two levels of detail in these calculations; 
firstly a `sampling' efficiency, in which we neglect stellar blending 
and  assume that all the additional flux in a microlensing event is
recovered by the photometry code; and secondly a `photometric' efficiency, 
where we add artificial stars to a representative set of real images, 
re-run the photometry code and create look-up tables of added 
vs.~recovered flux. 
These look-up tables are then used to generate artificial 
microlensing lightcurves in the same way as above. 
This `photometric' efficiency is more realistic, and is typically 
$\sim 20\%$ lower than the `sampling' efficiency for timescales less
than 200 days.  

The photometric efficiency is based on all stars in our fields, even
those not uniquely identified because of S/N or crowding effects.  These are
accounted for by integrating the detection efficiency per star over a
corrected luminosity function (LF) as in \yrone.  This corrected LF is
truncated about one magnitude beyond where our measured LF becomes
seriously incomplete.  However, the real LF continues to rise as
$10^{0.5m}$ for several magnitudes ($m$) beyond this cutoff so there should
be an additional contribution to our exposure from these fainter
stars.  We have tried several different magnitude cutoffs and it
appears that our exposure is converging near or below the cutoff used in
this paper for events with $\that < 150\,$days. 
Any additional contribution should be relatively
insignificant for reasons discussed in the appendix.

Efficiency results are shown in Figure~\ref{fig-eff}. 
We define our efficiency as the 
fraction of input events with $\umin < 1$ which pass our cuts; since
we use a cut of $\Amax > 1.75$ or $\umin < 0.661$, our 
efficiency is constrained to be less than 0.661. 
This efficiency is defined relative to an `exposure' of 
$E = 1.82 \ten{7}$ star-years, which arises as follows: 
there are 9.2 million lightcurves in our total sample, of which $8\%$ 
have insufficient valid data points to be used in the simulations, 
and 6.5\% occur in field overlaps. The relevant timespan 
is the 840-day interval over which we add the simulated events; 
thus the exposure is $7.9 \ten{6}$ stars $\times 840 \
{\rm days} = 1.82 \ten{7}$ star-years. 

\begin{figure}
\plotone{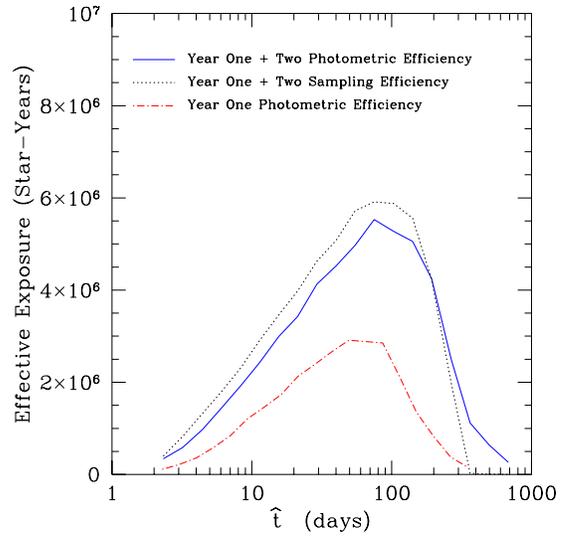}
\caption{`Effective exposure' $E \eff(\that)$ to microlensing 
 (normalised to $\umin < 1$) for the 2-year Macho data. 
 The solid line is the best 
 estimate using `photometric' efficiency; the dotted line
 is an upper limit using `sampling' efficiency.  For comparison,
 the dot-dash line shows the value for the 1-year analysis
 using the ``photometric" efficiency.  
   \label{fig-exp} } 
\end{figure}

The efficiency for timescales $\that \sim 10 - 60$ days
is lower than that from
\yrone\ by $\sim 10$\%; this is because we gain around 10\% from
loosening the goodness-of-fit and star quality cuts, but we lose about
20\% from the tighter $\Amax$ and $\Delta\chi^2$ cuts.  A more
operational comparison of the 1-year and 2-year samples is seen in
Figure~\ref{fig-exp} which shows `effective exposure' $E \eff(\that)$.  
 
\placefigure{fig-eff}
\placefigure{fig-exp} 
\begin{figure}
\epsscale{0.9}  
\plotone{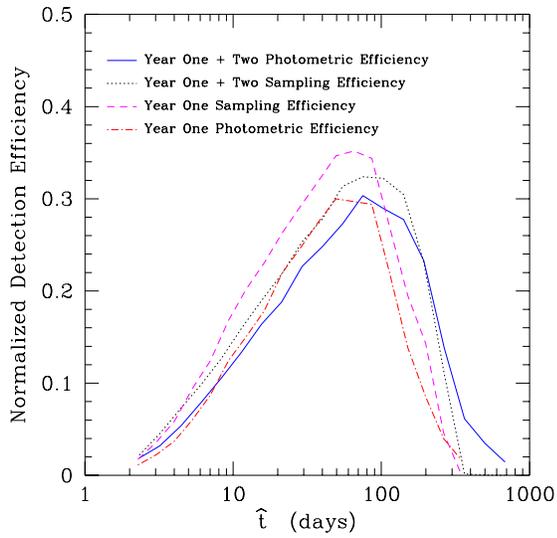}
\caption{ Microlensing detection efficiency (normalized to $\umin < 1$)
 for the 2-year Macho data, as a function of event timescale $\that$.
  The dotted line shows the `sampling' 
 efficiency, and the solid line shows the `photometric' efficiency as
 described in \S~\protect\ref{sec-eff}.   For comparison the
 corresponding curves from year-1 (\yrone) are shown.
   \label{fig-eff} } 
\end{figure}

The most substantial difference occurs at long timescales, where the
`new' efficiency is significantly higher than in \yrone.  This occurs
partly as a consequence of the longer timespan which allows longer
events to fit within our observing window; and partly because we have
relaxed the year-1 cut $\that < 250$ days, $t_{FWHM} < 100 $ days to
simply $\that < 300$ days.  The main motivation for the $t_{FWHM} <
100$ days cut was to give a second redundant cut, in addition to the
positional cut, to reject the clump of `nebulosity variables' in 30
Dor, which we earlier suspected to be T Tauri stars. These events have
now been recognized as arising from the light echo around SN 1987A
(c.f. \citeNP{xu95}); they seem to arise from small `knots' of
nebulosity illuminated by the light echo, hence the smearing of the
echo by the light-crossing time results in roughly symmetrical
brightening events for some of these objects.  Since we know that
there have been no other LMC supernovae in recent years, we can simply
reject all candidate events from this small patch of sky, and the
timescale cut can be relaxed.

There are a number of shortcomings of the present efficiency analysis.
One problem is that the luminosity function used for the simulated
events in the efficiency analysis does not go as deep as we would like. 
We are not very sensitive to short timescale events for very faint stars,
so our truncation of the luminosity function is not a problem, but for
longer timescale events we get significant detection probabilities at
faint magnitudes. For $\that > 200\,$days it appears that there may
be a significant contribution to the overall detection probability from
stars fainter than the ones we have included in our efficiency calculations,
so our `photometric' efficiency is probably an underestimate for these
long events. For detected events with $\that<75$days, we expect that the
systematic error in our 'photometric' efficiency due to the truncated
Monte Carlo LF is less than $5\%$.  For events with
$\that<150$ days this error is expected to be less than $10\%$
of our final efficiency values.

The efficiency determination also depends upon
how well the Monte Carlo simulations represent realistic distributions of
microlensing events. The main shortcoming of our Monte Carlo simulations
is that all events are assumed to be ``normal" microlensing events with
a single lens, a point source, and constant velocities. There are, of
course, observed microlensing events (such as event 9)
which violate these assumptions, and our detection efficiency estimates
are clearly overestimates of our actual efficiency of detecting these
exotic lensing events. This will cause our efficiencies to be overestimated
perhaps by as much as 5\%. (The factor may be larger than this for the
year-1 analysis which relied more heavily on fit $\chi^2$ cuts than 
the present analysis does.) 
Since two main shortcomings of our efficiencies have opposite signs and
magnitudes of 10\% or less, we feel that the total systematic error in our
efficiencies is probably less than 10\%. A somewhat more conservative
estimate of the systematic efficiency error is just the difference between
the photometric and sampling efficiencies which is about 15\% for events
with timescales in the observed range.

\clearpage
\section{ Event Distributions } 
\label{sec-dist}

There are a number of statistical 
tests that can be performed on microlensing event
distributions to test the hypothesis 
that our events are indeed gravitational microlensing, or to test
hypotheses regarding the lens population.
With only 8 events, these tests are not
conclusive, but they do support 
the interpretation that we have observed gravitational microlensing
by a new population.

\subsection{ Impact Parameters }
\label{sec-impact}

\begin{figure}
\plotfiddle{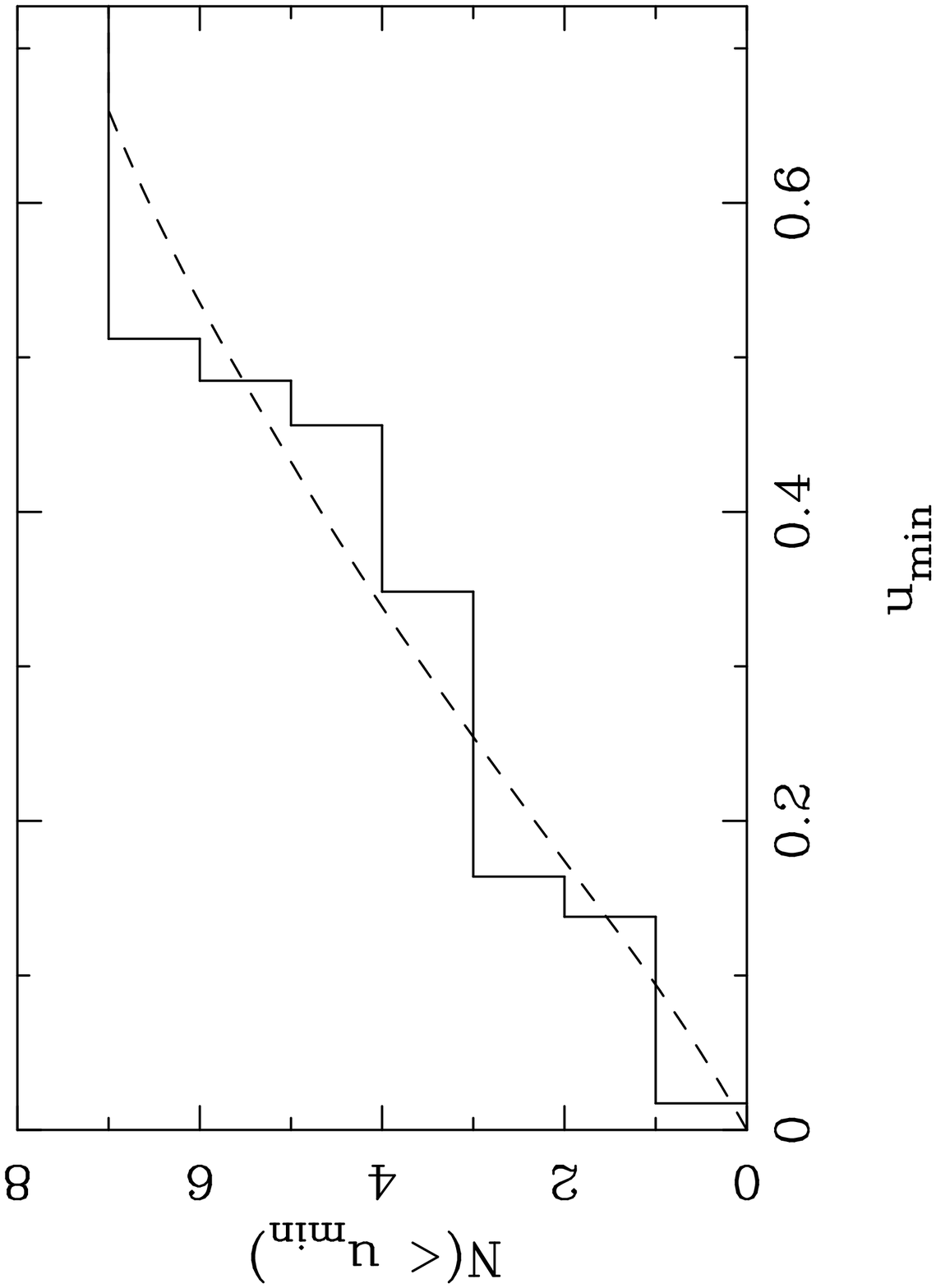}{3.0truein}{-90}{35}{35}{-130}{220} 
\caption{ The solid line shows the cumulative distribution 
 in $\umin$  of 7 candidate events (excluding the binary event~9). 
 The dashed line shows the expected distribution, i.e. a uniform
 distribution modified by our efficiency.  
   \label{fig-umin} } 
\end{figure}

An important model independent test of the
hypothesis that we have observed gravitational microlensing
is to compare
the distribution of peak magnifications to the theoretical prediction.
It is convenient to switch variables from the
maximum magnification ($A_{\rm max}$) to the minimum distance of approach
between the Macho and the line of sight, in units of the
Einstein radius,  $\umin = b / r_E$.
Events should be uniformly distributed in $\umin$; this distribution is
then modified by the experimental detection probability
which is higher for small
$\umin$ (high $A_{\rm max}$). 
This comparison was previously performed with a larger
sample of bulge events and the data were
found to be consistent with the microlensing hypothesis 
\cite{macho-bulge2}.
The observed and predicted distributions for our LMC events are shown in
Figure~\ref{fig-umin}; a KS test shows 
a probability of $96.6\%$ of getting a KS deviation worse
 than the observed value 0.177.
(The binary event~9 is excluded from this comparison because the
dependence of the detection efficiency
on $u_{\rm min}$ is substantially different for binary events.)
We conclude that the distribution of events in $\umin$
is consistent with the microlensing interpretation.

\placefigure{fig-umin}

The $\umin$ distribution and the high magnification events
may be used to lend support to the
interpretation of our lower magnification events\rlap.
The higher magnification events are striking,
and are clearly separated from the background in Figure~\ref{fig-cuts}.
They are distinct from any known type of intrinsic stellar 
variability, so they are very likely to be lensing events. Microlensing
predicts a uniform distribution in $\umin$ (with a slight modification
due to nonuniform detection efficiency), so we expect that (at least) 
some of the lower magnification events must also be caused microlensing.

This argument can be made quantitative.
Consider the event subsets \{1,5\}, \{1,5,7\}, and \{1,4,5,7\}.
These sets contain the highest 2, 3, and 4 magnification events respectively.
For each subset of $n$ events, we evalulate the mean
$\umin$ value: $\VEV{\umin}$. We then use a 
Monte-Carlo simulation using the efficiency-corrected $\umin$
distribution from Section~\ref{sec-eff} to 
compute the probability that a sample of n events will have 
a $\VEV{\umin}$ value as small as the observed value.
For example, $\VEV{\umin} = 0.080$ for events
1 \& 5, but only 3.2\% of simulated 2 event data sets have $\VEV{\umin}$
values this small. Thus, it is unlikely that events 1 and 5 are the 
{\it only} microlensing events which pass our cuts, so 
at least one of the other events should also be microlensing. 
Table~\ref{tab-umin} shows the results of this test for a number of
different data subsets. This table indicates that it is also unlikely
that the 3 or 4 highest magnification events are the {\it only} real
microlensing events which pass our cuts. When we consider event subsets
which include one or more of the events with $\Amax < 2.5$, then the
$\VEV{\umin}$ values are much closer to the expected value of
$\VEV{\umin} = 0.310$. In particular, the set of all non-binary microlensing
candidates, \{1, 4-8, 10\}, and the set of all non-binary events
classified as excellent or good microlensing candidates, \{1, 4-8\}, both
give quite acceptable values for $\VEV{\umin}$. If we consider the set
of non-binary 
`excellent' microlensing candidates, \{1, 4, 5\}, we find that their
$\VEV{\umin}$ is marginally incompatible (at the 90\% c.l.)
with the hypothesis that they are the only microlensing events to pass our
cuts. Thus, when we add in the binary event (which is classified as excellent)
we find that the $\VEV{\umin}$ test indicates that at least 5 of our
microlensing candidates are likely to be actual microlensing events. 
If the ``5th" real microlensing event is event 7, then the set of non-binary
candidates \{1, 4, 5, 7\} still have an unexpectedly small value of
$\VEV{\umin}$. Thus,
if the number of real lensing events is $< 6$, then it is probably the case
that event 7 which has a symmetric lightcurve with $\Amax \approx 6$
is a variable star of a previously unknown class. The microlensing
interpretation seems more likely than this.

\pagebreak
\subsection{ Color-Magnitude Diagram }
\label{sec-cmd}

\begin{figure*}
\epsscale{2.0}
\plotone{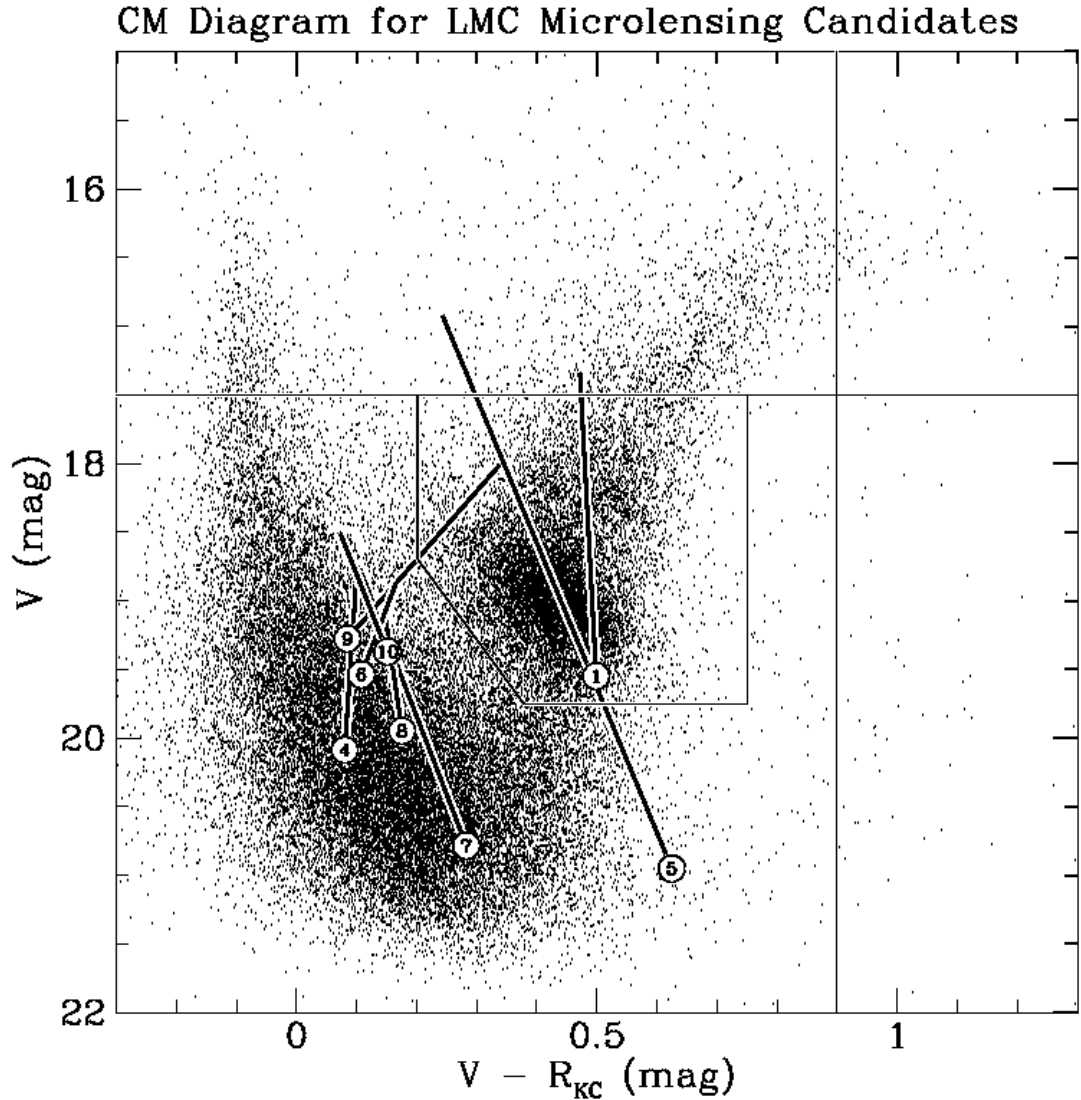}
\caption{The locations of the 8 candidate microlensing events on
 the color-magnitude diagram. The events at normal brightness 
 are shown as numbered circles,
 with bars indicating their locations at peak brightness. The small dots
 show all stars from a $5' \times 5'$ box around each event. The regions
 of the color-magnitude diagram which are excluded from the microlensing
 search ($V > 17.5$ and $V-R > 0.9$) are indicated with straight lines,
 and the ``giant branch region" is indicated by the closed figure.
 \label{fig-cmd} }
\end{figure*}

The gravitational microlens does not distinguish between types
of star. The stars which have undergone microlensing should
be ``democratically" distributed over the color magnitude
diagram for the LMC.
Figure~\ref{fig-cmd} shows a color magnitude
diagram with each of the 8 microlensing
candidates along with all the stars in a $5\pri \times 5\pri$ region
around each candidate.
\placefigure{fig-cmd}
Most of the events are along the faint main sequence
where most of the observed LMC stars reside. 
Using the Monte Carlo simulations that were run to determine our
microlensing event detection efficiencies, we
have computed the fraction of lensing events in which the source star
appears to be a clump giant star (the clump giants are
defined for this purpose by the box indicated in Figure~\ref{fig-cmd}).
For events with timescales of $\that\approx 75\,$days, 
this fraction is 10\% which is quite consistent with the 1 of our 8
events which resides in the clump.
Event 5 appears to lie in a sparsely
populated region of the diagram, but this is because the magnified
star (which lies on the main sequence) is blended with a much redder
object. We conclude that the distribution of events 
is not clustered in the color magnitude diagram and is consistent with
the microlensing interpretation.

\subsection{ Spatial Distribution }
\label{sec-spatial}

For microlensing by Machos smoothly distributed in the Galactic halo, we
expect the detected events to be distributed across
our fields in proportion to the local exposure, $E$.  
In contrast, models in which LMC objects dominate the lensing
population predict that the lensing events will be concentrated within the
LMC bar \cite{wu94,sahu94}. 

\begin{figure}
\epsscale{0.9}
\plotone{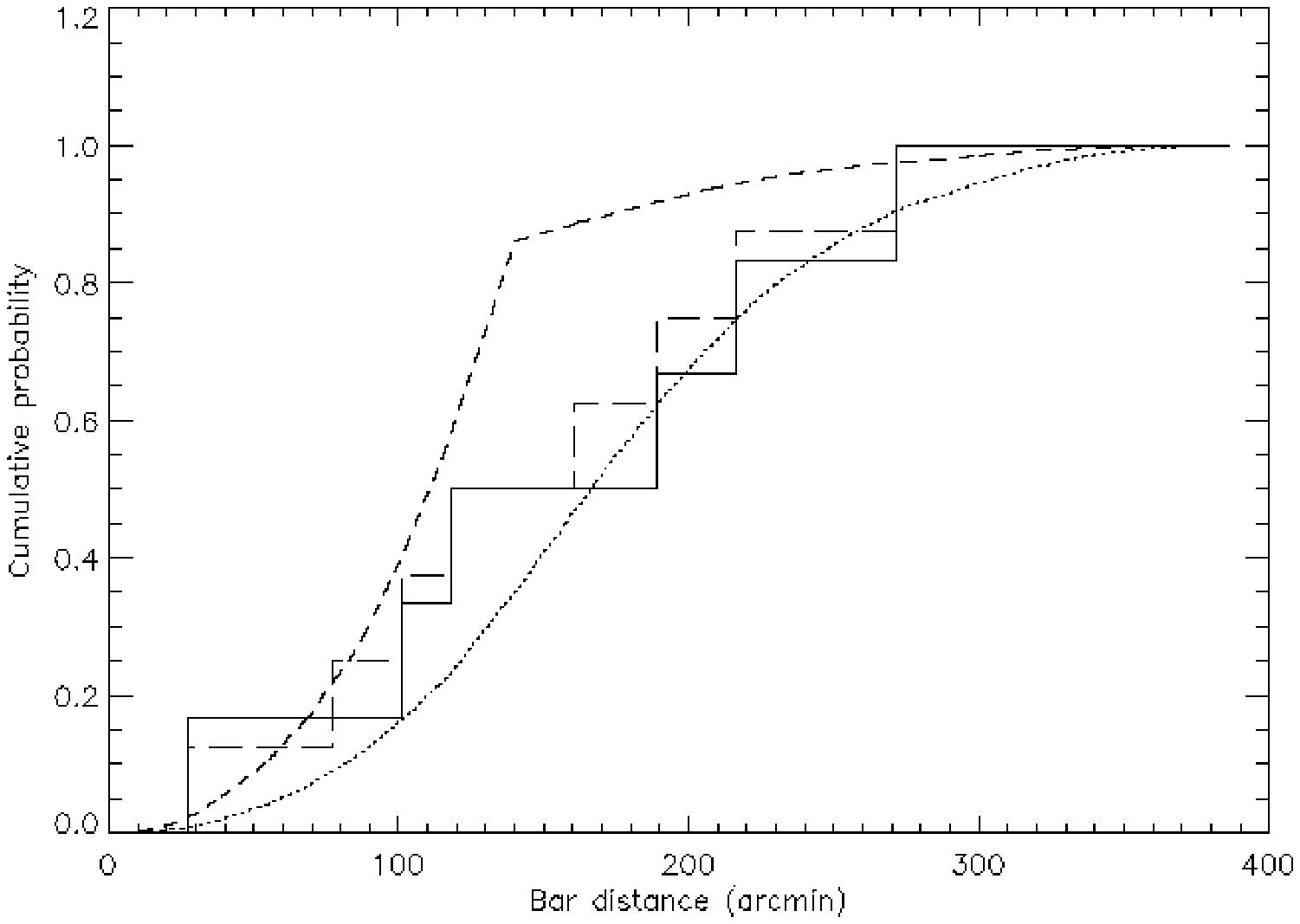} 
\caption{ Cumulative distributions of the ``distances" of the events from the
center of the LMC bar are shown for the 8 and 6 event samples (long dashed
and solid curves respectively) as well as for the ``pure halo lensing"
(dotted curve) and the extreme bar lensing models (short dashed curve).
   \label{fig-lmc_space} }
\end{figure}

Figure~\ref{fig-lmc} indicates that the detected
events are apparently spread evenly across our 22 fields. 
To quantify this impression, we consider a simple model for the LMC bar: 
the bar is an ellipse
with semi-major axis of 70 arcmin and axis ratio 4:1. We
define a ``distance" of a point from the center of the bar to be the
semi-major axis of a similar ellipse which passes through that point.
We consider the expected distribution of events over
this distance from the bar, under three models: (1) the
microlensing optical depth does not vary over our fields
(pure Galactic halo microlensing); (2) 
the microlensing optical depth is 4 times greater inside the bar
than outside (large contribution from LMC lenses); and (3) 
the microlensing optical depth is 12 times greater inside the bar
than outside (huge contribution from LMC lenses). Models (2) and
(3) bracket the models proposed by \citeN{sahu94} who suggested that
the microlensing optical depth inside the bar should be 4-12 times
larger than the optical depth outside the bar.
The cumulative distributions of bar distances resulting from 
Monte Carlo realizations of these three models are compared with 
our 8 and 6 events samples in 
Figure~\ref{fig-lmc_space}.

We use non-parametric statistical tests to
quantify the match of the observations to the models. Specifically,
we have used one- and two-sided Kolmorogov-Smirnov tests and
the Wilcoxon two sample test \cite{kendall-stuart}.
This latter test utilizes only 
the rank ordering of the data
and is especially powerful for testing whether one distribution is
systematically shifted from another.
As we apply this test, the null
hypothesis is that the distributions of bar distances
are identical between model and observation, while the alternative hypothesis
is that the bar distances for the observations are greater than those for
the model.  The probability of the null
hypothesis is given for
each model in Table~\ref{tab-space}.
We note that as this probability decreases,
the likelihood that the model is
correct decreases.

The results show that the most extreme LMC lens model is ruled out at roughly
a 90\% confidence level by the most powerful statistical tests, 
with the less extreme being ruled out at only about 70\%
confidence.  Clearly, this result is not yet conclusive, but with the
additional data that will be available within a few years, it should
be possible to determine whether or not LMC microlensing makes a significant
contribution to the total.

\clearpage
\section {Implications} 
\label{sec-implic}

In this section we examine the consequences of the observed
candidate microlensing events for the dark matter in the Milky Way.
We start with the inferred microlensing optical depth, which is compared
with the optical depth expected from a ``standard" dark halo comprised
entirely of Machos. We discuss the sensitivity of our conclusions
to the detection efficiencies, and to the selection criteria.
We then discuss the observed microlensing rate, and the durations of
the events we have seen.  We are able to obtain a plausible estimate
of the total mass in Machos, and to place a strong upper bound on the
contribution of low mass Machos.
We employ maximum likelihood estimators to infer typical masses and 
probable Macho fractions for various model halos.
Finally, we discuss the possibility that some or many of
our observed events arise from objects not in the dark halo of the 
Milky Way.

\subsection{ Optical Depth Estimates }
\label{sec-tau}

The simplest measurable quantity in a gravitational microlensing experiment is
the microlensing optical depth, which is defined to be the instantaneous
probability that a random star is magnified by a lensing object by more than
a factor of 1.34. This is related to the mass in microlensing objects
along the line of sight to the source stars by
\begin{equation} 
\label{eq-taudef}
 \tau = {4 \pi G \over c^2} \int \rho(l) {l (L-l) \over L} \, dl 
\end{equation} 
Thus, it depends only on the density profile of lenses, not on their
 masses or velocities.  
Experimentally, one can
obtain an unbiased estimate of the optical depth as 
\begin{equation}
\label{eq-taumeas}
 \tau_{\rm meas} = {1 \over E} {\pi\over 4} 
                     \sum_i {\hat t_i \over \eff(\that_i)} \ .
\end{equation}
where $E$ is the total exposure (in star-years), $\that_i$ is the
Einstein ring diameter crossing time, and $\eff({\that_i})$ is the
detection efficiency. 
Here, and below, we use the average blend corrected values of 
$\that_{bl}$
from column 3 of Table~\ref{tab-blthat}.  
These take into account the fact that our typical
star is blended, and so the fit $\that$ is typically underestimated. 
As above, our total exposure is $E=1.82 \ten{7}$ star years.

It is also convenient to define the function 
\begin{equation}
\label{eq-tau1}
 \tau_1(\that) = {1 \over E} {\pi\over 4} 
                    {\hat t \over \eff(\that)} \ , 
\end{equation}
which is the contribution to $\tau_{\rm meas}$ 
from a single observed event with timescale $\that$. $\tau_1$ values 
for each of our events are also listed in Table~\ref{tab-blthat}.
Confidence levels
on our measured value of $\tau_{\rm meas}$ are determined with the
following procedure. We consider sets of events chosen according to
Poisson statistics with event timescales randomly selected from the
observed set of timescales. The Poisson distribution has a parameter, the
number of events expected ($N_e$), which is adjusted so that some fixed
fraction of the generated data sets (say, 16\%) have `measured' optical depths
that are larger than our actual value of $\tau_{\rm meas}$. The mean 
optical depth for the selected value of $N_e$ is then selected as the 
confidence limit value. (In our example, the confidence level would be
16\%--or a 1-$\sigma$ lower limit on $\tau_{\rm meas}$.) Our results
for $\tau_{\rm meas}$ and confidence intervals are shown in 
Table~\ref{tab-tau}. As explained below,
this procedure can underestimate the
errors when events much longer than those detected are not highly unlikely,
but it gives a reasonable estimate for the lower limit on $\tau_{\rm meas}$.

Using our full sample of 8 events, we find an optical depth 
for events of duration $2$ days $< \that < 200$ days of
$\tau_2^{200} = 2.9 {+1.4 \atop -0.9} \ten{-7}$. If we subtract the
predicted background microlensing optical depth of
$\tau_{\rm backgnd} = 0.5 \ten{-7}$ (from Table~\ref{tab-stars} below), we
find that the observed excess is about 50\%
of the predicted microlensing optical depth for a ``standard''
 all-Macho  halo of equation~\ref{eq-stdhalo} below. Alternatively,
we can estimate the optical depth due only to the halo 
by using the 6 event subsample defined in 
section \ref{sec-events}, for which   
$\tau_{2}^{200} = 2.1 {+1.1 \atop -0.7} \ten{-7}$,  
about 45\% of the
optical depth predicted by a ``standard" all-Macho halo.

\begin{figure}
\plotfiddle{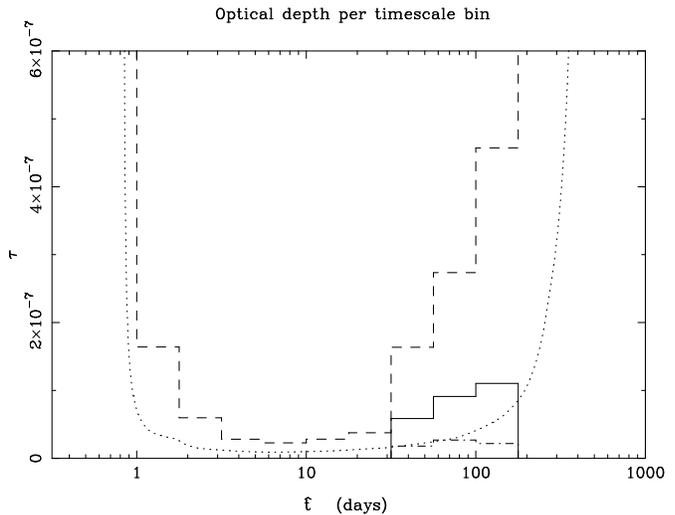}{3.0truein}{-90}{36}{36}{-150}{250} 
\caption{ The contribution to the optical depth 
 of eq.~\protect\ref{eq-taumeas}  from 
 events binned in timescale. The solid line shows the observed 
 values from the 8-event sample; the dashed and dot-dash lines show the 
 90\% confidence upper and lower limits (see text). 
 The dotted curve shows the contribution to $\tau$ which would arise 
 from a single observed event with timescale $\that$. 
   \label{fig-tau} } 
\end{figure}

This optical depth estimate has the virtue of simplicity; 
 however, since the events are ``weighted'' $\propto \tau_1$, 
 it is hard to assign meaningful confidence intervals 
 to $\tau$ without assuming some particular $\that$ distribution. 
This is illustrated in Figure~\ref{fig-tau}, which shows the 
 contribution to the sum in eq.~\ref{eq-taumeas} 
 from events in various bins  of $\that$. 
 Within each individual bin, the events have 
 similar `weights' and the uncertainties are well approximated by 
 Poisson statistics. 
 The confidence intervals in Fig.~\ref{fig-tau} are derived
 as follows: for each bin, we derive upper and lower limits $N_{\rm up}, 
 N_{\rm lo}$ 
 on the expected number of events in the bin in the usual 
 way, from the observed number of events in the bin  
 and Poisson statistics. 
 Then, we evaluate the maximum and minimum contribution 
 to $\tau_{\rm meas}$ from a single hypothetical event 
 in the timescale bin, 
 i.e. $\tau_1^{\rm min} = {\rm min}( \tau_1(\that); \that \in {\rm bin}) $
 and likewise for $\tau_1^{\rm max}$. 
 This gives limits $\tau_{\rm lo} = N_{\rm lo} \tau_1^{\rm min}$ and 
 $\tau_{\rm up} = N_{\rm up} \tau_1^{\rm max}$ for each bin, 
 which are actually  somewhat conservative.  

 This figure illustrates two important points: 
 firstly, the absence of short events with $2 < \that < 30$ days
 places strong upper limits on the optical depth from events in 
 this interval. 
 Although the efficiency is smaller here, an observed event 
 would contribute 
 $\tau_1(\that) \propto \that_i / \eff(\that_i)$ which is small. 
 Thus, the effective ``sensitivity'' of the experiment, in terms
 of number of detected events per unit optical depth, is proportional to 
 $\eff(\that)/\that$, {\bf not} just $\eff(\that)$. 
 This function is maximal at around $\that \sim 7$ days. 
 
Secondly, the uncertainties become very large at longer 
 timescales, due to the combination of 
 increasing $\that$ and decreasing $\eff(\that)$. 
  Thus the  overall uncertainty in 
 $\tau$ is considerably greater than simple 
 Poisson statistics based on 6 or 8 events. 
For example, if we should expect to have observed on average 
 1 additional event with  $\that \sim 300$ days, 
 but we happened to observe no such event, 
  the real $\tau$ would be $2 \times$ the above estimate, and 
 we clearly cannot exclude such a possibility with any confidence. 
Thus, it is dangerous to quote any optical depth result as an `upper limit on 
 Machos' without specifying a mass or timescale 
 interval to which this limit applies. 

\subsection{ Optical Depth Efficiency Dependence }
\label{sec-tau_eff}

Table~\ref{tab-tau} includes a number of estimates in addition to our
best estimate of the optical depth confidence levels, which is given
in bold face type and uses the
photometric efficiency with the average $\that$ corrections.
While the photometric efficiency is the best estimate of our detection
efficiency, it is a slight underestimate of our detection efficiency for 
`normal' (single lens, constant velocity) events because we have not
included the contribution due to very faint stars as explained above.
This leads toward a slight underestimate of $\eff$, while our efficiency
for detecting exotic microlensing events has been overestimated.
Thus, there are two known systematic efficiency errors which tend to
cancel. To get some idea how much these problems might affect our optical
depth calculations, we have calculated $\tau_{\rm meas}$ using the sampling
efficiencies which are expected to be overestimates of the true detection
efficiencies for events in the range 
$20\,{\rm days} < \that < 150\,{\rm days}$. Thus, we expect that the 
sampling efficiency $\tau_{\em meas}$ confidence levels given in 
Table~\ref{tab-tau} are likely to be underestimates of the true microlensing
optical depth.

Another variation of our $\tau_{\rm meas}$ values that are shown in
Table~\ref{tab-tau} is due to our choice between $\that$ correction methods
due to blending. We have used the average $\that$ corrections given in
Table~\ref{tab-blthat} for our `official' results, but another choice would be
to use the blend fit values given in Table~\ref{tab-blend}. The blend fit
values are certainly more accurate on an event-by-event basis, but we are
concerned that they might tend to systematically overestimate the true
$\that$ value. As shown in Table~\ref{tab-tau}, this choice has
little effect on our results.

Also included in Table~\ref{tab-tau} are two lines using sampling efficiencies
with {\it no} $\that$ correction at all. This is likely to give rise to
a substantial underestimate of $\tau_{\rm meas}$, but we include it because
(as discussed above) it gives a firm lower limit on $\tau_{\rm meas}$.
A reasonable estimate of our 1-$\sigma$ systematic error in our efficiency
determination is given by the scatter of all the $\that$ corrected values
listed in Table~\ref{tab-tau}. This is about 10\% which is much smaller than
the statistical errors.

The last two rows of Table~\ref{tab-tau} give confidence limits on $\tau$
for the 6 event subsample. These are included to allow the reader to
assess the effect of removing some of the selected events from the
sample. Event 10, for example, has a lightcurve asymmetry that 
resembles that of a variable star. If we remove this event, and the longest
event (event~9), we have our 6 event sample. Table~\ref{tab-tau} indicates 
that even with 6 event sample and the upper limit efficiencies 
(\ie, sampling efficiencies with no $\that$ correction), the 97.5\% 
confidence level lower limit on the microlensing optical depth is still
larger than total microlensing optical depth of 
$\tau_{\rm backgnd} = 0.54 \ten{-7}$ given in Table~\ref{tab-stars} below.
Thus, the excess microlensing optical depth over the background prediction
is not sensitive to the uncertainties in our microlensing detection
efficiencies or to the possibility that a small fraction of our candidate
microlensing events might actually be variable stars.

\subsection{ Optical Depth Cut Dependence }
\label{sec-tau_cut}
\begin{figure}
\plotone{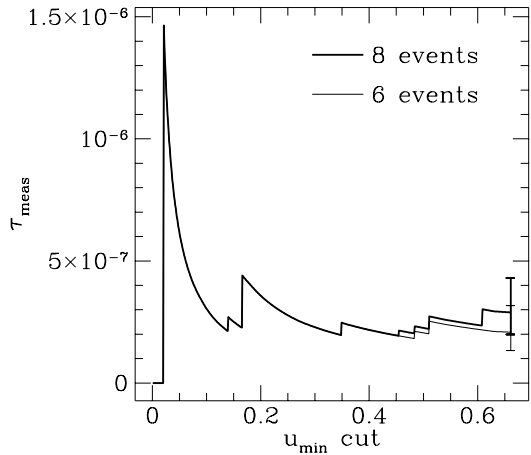}
\caption{ The measured microlensing optical depth is plotted
 as a function of the $\umin$ cut for the 8 event sample (thick line)
 and the 6 event `halo' sample (thin line). The error bars are 
 indicated at our selected cut $\umin \leq 0.661$  ($\Amax \geq 1.75$.)
   \label{fig-tau_ucut} } 
\end{figure}

\begin{figure}
\plotone{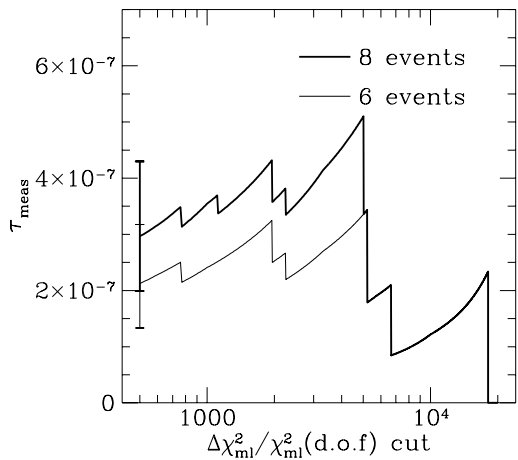}
\caption{ The measured microlensing optical depth is plotted
 as a function of the $\Delta\chi^2/(\chi^2_{ml}/\ndf)$
 cut for the 8 event sample (thick line)
 and the 6 event `halo' sample (thin line). The error bars are 
 indicated at our selected cut 
 $\Delta\chi^2/(\chi^2_{ml}/\ndf) \geq 500$.
   \label{fig-tau_delc2cut} } 
\end{figure}

Figures~\ref{fig-tau_ucut} and \ref{fig-tau_delc2cut} show the dependence
of the measured optical depth on the $\umin$ and 
$\Delta\chi^2/(\chi^2_{ml}/\ndf)$
cuts. The heavy curves indicate $\tau_{\rm meas}$ for the full 8 event
sample while the light curves give $\tau_{\rm meas}$ for the 6 event
subsample. For the binary event we have assigned a $\umin$ value of 0.609
which is the value obtained for the single lens fit. Another option would
have been to use the binary fit value of $|\umin | = 0.054$. This is not
appropriate, however, because the binary fit includes blending whereas our
cut used the unblended single lens fit value. Strictly speaking, the single
lens fit value would be correct if we had used a binary event detection
efficiency in order to calculate the binary event's contribution to
to $\tau_{\rm meas}$. 

We have also adjusted the $\Delta\chi^2/(\chi^2_{ml}/\ndf)$ 
values used in
Figure~\ref{fig-tau_delc2cut} for the binary event and the alert event
(event~4). For the binary event $\Delta\chi^2 = 7191$ while 
$\chi^2/\ndf = 6.868$ for the single lens fit and
$\chi^2/\ndf = 1.755$ for the binary lens fit. For 
Figure~\ref{fig-tau_delc2cut} we have used the binary fit value of
$\chi^2/\ndf$ implying
$\Delta\chi^2/\chi^2_{ml}/\ndf = 5018$ for the binary event.
The alert event was detected in progress substantially before peak
magnification, and this resulted us obtaining more observations than
normal of field 13 (which includes this star). We have taken up to
4 observations of field 13 per night when we would normally have taken
only 1 or 2. Thus, for event~4, 
$\Delta\chi^2/(\chi^2_{ml}/\ndf) = 6728$ is larger
than we would get with our normal observing strategy by a factor of
$\sim 3$. Therefore, we have reduced the value of 
$\Delta\chi^2/(\chi^2_{ml}/\ndf)$ used for event~4 in
Figure~\ref{fig-tau_delc2cut} to 2243.

Figures~\ref{fig-tau_ucut} and \ref{fig-tau_delc2cut} clearly indicate
that our optical depth results are not very sensitive to our choices
of cut values. The $\tau_{\rm meas}$ values generally do not vary by
more than the 1-$\sigma$ statistical error bars for $\umin$ and
$\Delta\chi^2/(\chi^2_{ml}/\ndf)$ cuts in the ranges
$0.1 \leq \umin \leq 0.661$ and
$ 500 \leq \Delta\chi^2/(\chi^2_{ml}/\ndf) \leq 5000$.

\clearpage
\subsection{Microlensing Rate}
\label{sec-rate}

 The event rate $\Gamma$ is more model-dependent than 
 the optical depth $\tau$, since 
 it depends on the event timescales 
 via the mass function of Machos and 
 their velocity distribution, but the uncertainties on 
 $\Gamma$ are given purely by 
 Poisson statistics and thus can give very robust limits 
 once the halo model is specified. 

 The number of observed events 
is given by Poisson statistics with a mean of 
\begin{equation} 
\label{eq-nexp} 
\Nexp = E \int_0^{\infty} \, {d\Gamma \over d\that} \,
    \eff(\that) \, d\that
\end{equation} 
where $ E = 1.82 \ten{7}$ star yr is our total `exposure'.  
As in \yrone, we have compared our results with the commonly used
 model of the dark halo (hereafter model S) 
\begin{equation}
\label{eq-stdhalo}
 \rho_H(r) = \rho_0 { R_0^2 + a^2 \over r^2 + a^2 } 
\end{equation} 
 where $\rho_H$ is the halo density, 
 $\rho_0 = 0.0079 \msun \, \pc^{-3} $ is the 
 local dark matter density, $r$ is Galactocentric radius, 
 $R_0 = 8.5 \kpc$ is the Galactocentric
 radius of the Sun, and $a = 5 \kpc$ is the halo core radius.  
We assume a Maxwellian distribution of Macho velocities
with an 1-D rms velocity of $155 \kms$, 
and (initially) assume a delta-function Macho mass function
of arbitrary mass $m$. The resulting microlensing rate 
$d\Gamma/d\that$  is given by equation~(A2) of \yrone, 
and the total rate is 
$\Gamma = 1.6 \ten{-6} (m/\msun)^{-0.5}$ events/star/yr.  
Thus if our efficiency were 100\%, we would expect 
about $30 (m/\msun)^{-0.5}$ events in the present data set; 
note that this implies over 1000 events if the halo is made 
of Jupiters, but only 10 (of longer duration) 
if it were made of $10 \msun$ black holes;  
thus, although the optical depth is independent of Macho mass, 
in a real microlensing experiment 
the limits on the Macho content of the halo  
will be strongly dependent on the Macho mass.

\begin{figure}
\plotone{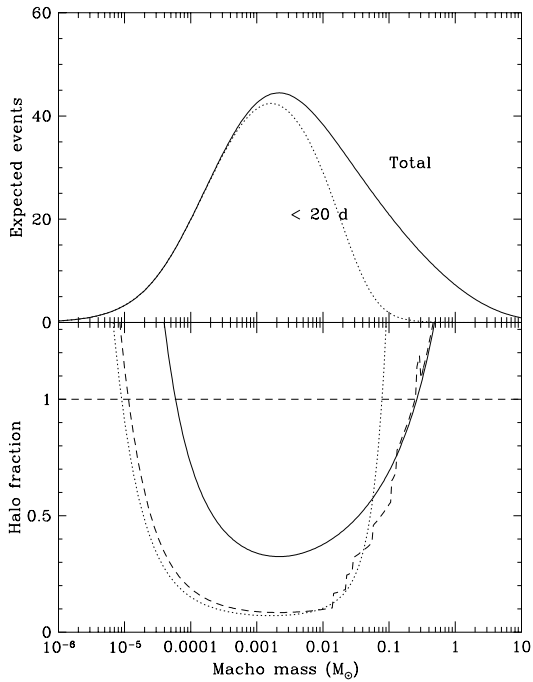}
\caption{ Upper panel: expected 
 events for halo model S, for an all-Macho halo 
 with delta-function Macho mass function. 
 The solid line shows the expected total number, the dotted line
 shows expected number of events with $\that < 20$ days. 
 Lower panel: Upper limits on Macho fraction of halo model S. 
 Regions above the curves are excluded at 95\% CL.
 The solid line is derived from 8 observed events, the dotted line 
 from 0 observed events with $\that < 20$ days, and the short-dash line 
 from the `80\% interval' analysis of \S\protect\ref{sec-lowmass}. 
   \label{fig-flim} } 
\end{figure}

 It is convenient to define $\Ntil(m)$ to be the number of expected
 events for an all-Macho halo and unique Macho mass $m$ ; this function 
 is shown in the upper panel of Figure~\ref{fig-flim}.
 It is roughly 1.8 times higher
 than the corresponding curve in \yrone, due to the doubling in 
 the exposure and the slightly reduced efficiency for the new 
 selection cuts.  As before, there are two competing effects; 
 for Macho masses $\gtrsim 0.01 \msun$, 
 most events have timescales $\that \gtrsim 10$ days where 
 our efficiency is 
 quite good,  but the event  rate is falling $\propto m^{-0.5}$. 
 For small masses $m \lesssim 10^{-3} \msun$, the theoretical event rate is 
 very high but most events are shorter than 
 $\that \sim 3$ days where our efficiency is very low. 
The product of these two effects causes the peak in $\Ntil(m) \approx 45$
 at $m \sim 10^{-3} \msun$. 

\placefigure{fig-flim}

Assuming that we have detected not more than 8 microlensing events 
passing our selection cuts, Poisson statistics exclude at $95\%$ 
confidence level any model which predicts a mean number of events 
$\Nexp > 14.5$; i.e.~we may obtain an upper limit on the 
fraction of the halo made of Machos of mass $m$, 
$\flim(m) = 14.5 / \Ntil(m)$; this is shown in 
the lower panel of Figure~\ref{fig-flim}.  
This limit also applies to arbitrary mass functions
within an appropriate interval; 
if we define $\psi(m) dm $ to be 
the fraction of halo density comprised of Machos in the mass interval
$[m, m+dm]$; then we have 
\begin{equation}
\Nexp = \int_0^\infty \psi(m) \Ntil(m) \,dm ; 
\end{equation}
thus as pointed out by Griest (1991) and \yrone, 
if we set a limit $f < f_0$ for some constant $f_0$ for 
a $\delta$-function Macho mass function over some
mass interval $m_1 < m < m_2$, we also limit the total halo  
fraction contained within the same interval for any mass function, 
i.e. $\int_{m_1}^{m_2} \psi(m) \, dm < f_0$; 
thus if we draw a line at $f_0 = 0.5$, we see from Figure~\ref{fig-flim}
that Machos from $2 \ten{-4}$ to $0.04 \msun$ contribute
less than 50\% of our standard halo. 
(We derive a stronger limit below using the event timescales). 

We can usefully constrain the total mass of Machos interior to $50 \kpc$.
The above model halo has a mass of $4.1 \ten{11} \msun$ 
within $50 \kpc$.
Thus, the limit on the Macho fraction may be expressed as 
a limit on the total mass of Machos interior to $50 \kpc$, 
$\Mlim(m) = 4.1 \ten{11} \msun (14.5 / \Ntil(m))$. 
This latter formulation is very useful, because \yrone\ showed 
that the limits on the Macho {\it fraction} are quite sensitive 
to the choice of halo model, but the limits on the {\it total mass} 
of Machos within $50 \kpc$ are much less sensitive.

\subsection{Limits on Low-Mass Machos} 
\label{sec-lowmass}

The above analysis utilized only the observed number of events 
$\Nobs$. However, there is additional information available in the
set of event timescales $\{ \that_i \}$.  
For the halo of equation~\ref{eq-stdhalo} 
the expected timescale of microlensing 
events depends on the mass of the lens as 
$\VEV{\that} \approx 140 \sqrt{m / \msun} {\rm days}$ 
\cite{griest91}. 
The fact that all 8 of our candidate events have 
$\that \ge 34$ days means that they are unlikely to result
from low-mass Machos with $m \lesssim 0.01 \msun$; thus, 
we can obtain considerably stronger limits 
on such objects by using this information. 

We have approached this in two complementary ways: 
firstly, we can simply note that we have no candidate event with 
$\that < 20$ days (where 20 days is an arbitrary cut
conservatively smaller than 34 days);  so we can define $\Nshort$ to be 
the expected number of events shorter than 20 days, and 
exclude at 95\% CL any halo model 
which predicts $\Nshort > 3.0 $ . 
\placefigure{fig-nshort}
This limit is shown as the dotted line in Figure~\ref{fig-flim}; 
for low-mass Machos it is 
much more restrictive than the limit in \S\ref{sec-rate}, 
since for Macho masses $< 0.01 \msun$ nearly all the events
are predicted to have $\that < 20$ days, 
so $\Nshort \approx \Nexp$, and the resulting limit on Machos 
becomes stronger by a factor of $3.0 / 14.5 \approx 0.2$. 
  
While the above approach is very simple, it is 
subjective,  since our choice of the 20-day cut is 
made {\it a posteriori}. 
We have therefore added  an alternative approach in which the 
timescale cuts are defined objectively for each Macho mass 
 using the theoretical distributions. 
In detail, for each Macho mass, we evaluate the cumulative 
distribution of expected timescales, 
\begin{eqnarray}
  N(\that;m) & = E \int_0^{\that} \, {d\Gamma \over d\that} (m) \,
    \eff(\that) \, d\that   \\  
  G(\that;m) & = N(\that;m) / N(\infty;m)
\end{eqnarray}  
i.e.~$G(\that;m)$ is the average 
fraction of observed events shorter than $\that$ for mass $m$. 
We then compute $t_1(m)$, $t_2(m)$ such that 
$G(t_1;m) = 0.1$, $G(t_2;m) = 0.9$; 
i.e.~an interval such that for Macho 
mass $m$, $80\%$ of detected events would have $t_1 < \that < t_2$. 
Then we count the observed number of 
events in this timescale interval, which we call $N_{80}(m)$, 
and compute the corresponding Poisson upper limit $L_{80}(m)$. 
This gives a limit on the halo fraction 
$ f_{\rm lim,80}(m) = L_{80}(m) / 0.8 \Ntil(m)$, 
and we repeat this procedure for each value of $m$. 

For example, given a Macho mass of $m = 0.1 \msun$, 
we find that $t_1 = 20$ days and $t_2 = 79$ days. 
We have $N_{80}(m) = 4$ observed events in this interval, so 
$L_{80}(m) = 9.2$ using Poisson statistics. We would 
expect $0.8 \Ntil(0.1 \msun) = 16.6$ events in this interval. 
Thus $f_{\rm lim,80}(0.1 \msun) = 0.55$. 

This limit is shown as the dashed line in Figure~\ref{fig-flim}; 
the `steps' in the curve occur as events move in and out of the
 80\% timescale window. 
We see that this limit is quite similar to the limit 
derived from $\Nshort < 3$ 
for low masses, and approaches the limit from $\Nexp < 14.5$ 
 for high masses, 
as expected since our observed timescales are most
 consistent with high Macho masses $\simgt 0.1 \msun$. 
The disadvantage to this approach is that the argument of 
\yrone\ extending the limits to a continuous mass function is 
no longer valid, since the ``observed'' number of events 
is now a function of the assumed $m$. 
However, this analysis is useful in that it confirms the 
short-timescale limit  from $\Nshort < 3$  in an objective way. 

We conclude that Machos in the mass interval $8 \ten{-5} - 0.03 \msun$
contribute less than 20\% of the `standard' halo; 
we extend this to other halo models in \S~\ref{sec-models}. 
In a companion paper \cite{macho-spike}, we apply a complementary 
search for very short events 
(0.1 ${\rm day} < \that < 3$ days) to the current data set; 
after applying suitable selection criteria, 
we find no candidates, and thus we 
extend these limits to lower masses $\sim 3 \ten{-7} \msun$.

\subsection{Likelihood Analysis and Lens Masses} 
\label{sec-like}

In the above section, we have presented upper limits 
on low-mass Machos which are valid regardless of the cause of 
our detected events. 
However, given that we have a substantial number of events, 
and the optical depth is a significant fraction of that 
 expected from an all-Macho halo, 
we need to assess the implications if our observed
candidates do result from microlensing by halo objects. 
Unlike the previous limits on Machos,  
the conclusions of the next 2 sub-sections depend on the assumption
that all 6 or 8 events are due to microlensing by objects in the halo.
As discussed in \S~\ref{sec-stars}, the expected 
microlensing contribution from all
non-halo populations is about 1 event.

For simplicity, we perform likelihood analyses based upon the 6 and 8
event samples similar to those done in \yrone.  A more detailed analysis
that includes in the likelihood function models for the density and
velocity distribution of background lenses is in progress and will
be published elsewhere.  As discussed in Section~\ref{sec-nevents},
the 6 event sample is a reasonable choice for a ``halo only" subsample.

\begin{figure}
\epsscale{0.9} 
\plotone{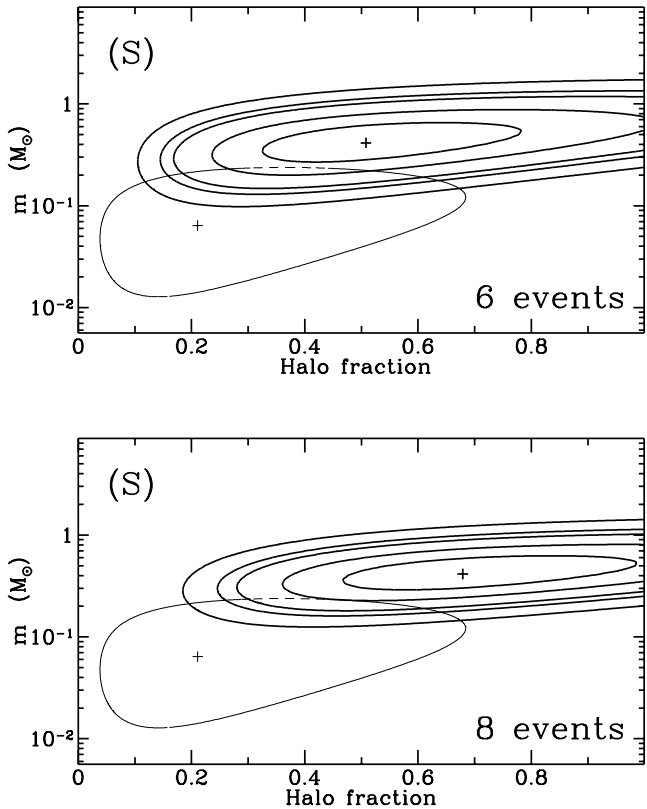}
\caption{ Likelihood contours of Macho mass $m$ and halo 
 Macho fraction $f$ for a delta-function mass distribution, 
 for halo model S. 
 The most likely value is indicated with a $+$, and the contours
 enclose total probabilities of $34\%, 68\%, 90\%, 95\%, 99\%$, 
 using a Bayesian method as described in \S~\protect\ref{sec-like}. 
 The light line shows the $90\%$ contour from \yrone. 
   \label{fig-like} } 
\end{figure}

Since the timescale of a lensing event is proportional to $\sqrt{m}$, 
we may use the observed timescales to estimate the lens masses, 
using a maximum-likelihood method as in \yrone. 
We use a 2-parameter model where a fraction $f$ of the 
 dark halo is made of Machos with a unique mass $m$ (the remaining $1-f$
 of the halo is assumed to be in Machos outside our mass range, 
 or particle dark matter). 
The likelihood of finding a set of $\Nobs$ detected events
with timescales $\that_i, i=1, \ldots, \Nobs$ is given by 
\begin{equation}
\label{eq-like}
  L(m,f) = \exp(-f\Ntil(m)) \prod_{i=1}^{\Nobs} \left( f E \eff(\that_i) 
  {d\Gamma \over d\that}(\that_i;m) \right), 
\end{equation}
where $d\Gamma/d\that$ is the theoretical rate of microlensing derived from
a halo model.  The results are dependent on the model so 
in \S~\ref{sec-models} we explore a range of possible halos.  For model S, 
the resulting likelihood contours, assuming a delta-function mass
function, are shown in Figure~\ref{fig-like}; the probabilities
are computed using a Bayesian method with a prior uniform in 
$f$ and $\log m$. 
\placefigure{fig-like}
We show plots for both our 6 event sample and our 8 event sample.  
We think the the 6 event sample gives the more reliable estimate
for halo lenses, but we also show the results of the 8
event sample to span the range of possibilities.
The peak of the likelihood contours gives the most probable mass and halo
fraction for this model and for the 6 event sample we find 
$m_{2D} = 0.41 \msun$, and $f_{2D} = 0.51$. 

We calculate the one-dimensional likelihood function by integrating over
the other parameter and find (for the 6 event sample)
a most likely Macho mass $m_{ML}= 0.46^{+0.30}_{-0.17} \msun$,
and most likely halo fraction $f_{ML} = 0.50^{+0.30}_{-0.20}$.
The errors given are 68\% CL.
The values for the 8 event sample are also given in Table~\ref{tab-like}.
It is important to note the large extent of the contours in 
Figure~\ref{fig-like}.  
This is mostly due to the small number of events.  
For model S, the 95\% CL contour includes halo fractions from 17\% to 100\%,
and Machos masses from 0.12 to 1.2$\msun$.

The most probable mass and halo fraction are both larger 
 than our results from year-1, which were $m = 0.06 \msun$ and $f = 0.2$, 
 though there is a reasonable degree of overlap of the contours.  
 The year-1 90\% CL contour is shown as the light line in
Figure~\ref{fig-like}.
The most probable value of each analysis lies outside the 90\% CL contour
of the other analysis.
 This is due to the fact that events 2 \& 3 from \yrone\ (with 
 $\that = 20$ and 28 days) 
 have dropped out of our new sample, 
 and all of the new events are longer than all the \yrone\ events. 
 This probably reflects small-number statistics, and the 
 improved efficiency for longer-timescale events with the 2-year data.

 A notable feature with the new results is that a delta-function 
 mass of $0.1 \msun$ is marginally excluded. This would 
 exclude brown dwarfs as the dominant lens population, 
 since a realistic brown-dwarf population 
 cannot include a significant fraction of objects above 
 $0.1 \msun$ due to star-count data (e.g.  \citeNP{bfgk,hhgc}), 
 while it should include smaller masses which would worsen the likelihood. 

 Given the importance of such a conclusion, we next ask how dependent are the
 results obtained here to the halo model and to delta-function mass 
 distribution assumption.

\subsection{Power-Law Halo Models} 
\label{sec-models} 

Although the halo model of eq.~\ref{eq-stdhalo} is a widely used baseline
 for comparison, there are few tracers of the mass distribution in the
 outer Galaxy and thus a considerable uncertainty in the real halo 
 parameters. 
To check the model dependence of our results, 
we have repeated our analysis for a wide
 range of `power-law' halo models \cite{evans93}. 
For simplicity,
we have used the same set of models as in Table~2 of \yrone, and
they are described in detail there. The set of models used span
an extreme set of probable halo masses, so any conclusions which hold
for this set of models are probably quite robust.  Briefly,
the models consist of extremes which match the measured Milky Way
rotation curve at $r_0$ and $2r_0$, and which have a range of disk
masses.  They do not include the Galactic bulge or bar.
Model A is a power-law equivalent of model~S; 
model B has a very massive halo with rising rotation curve; 
model~C has a light halo with a falling rotation curve, 
and model~D has the same rotation curve as model~A but a halo flattened to
about E6. 
Models~E, F and G have more massive disks and less massive halos,
with model~E being an extreme `maximal disk/minimal halo' model 
and models F, G being more realistic `heavy disk' models.  

\begin{figure}
\plotone{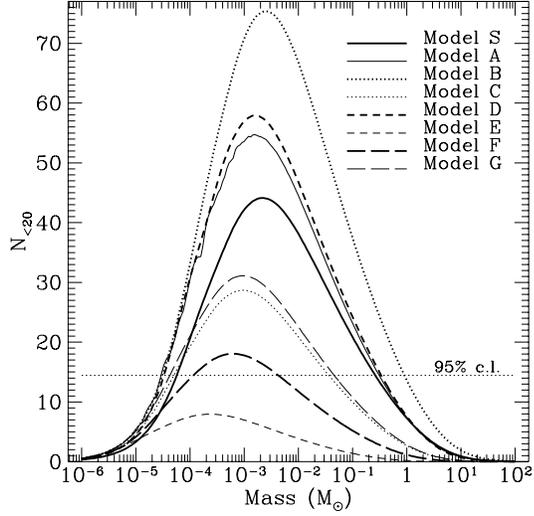}
\caption{ Expected number of events for the 8 halo models
 S, A-G,  with all-Macho halo and delta-function mass function. 
 The 95\% CL upper limit is for the 8 event sample.
   \label{fig-nexpall_nocut} } 
\end{figure}

\begin{figure}
\plotone{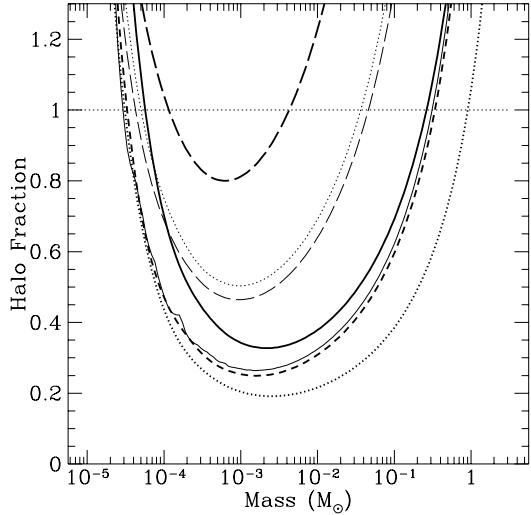}
\caption{ Upper limits (95\% CL) on the fraction of the halo made of 
 Machos for the 8 halo models S, A-G based upon 8 observed events.
 Line coding as Figure~\protect\ref{fig-nexpall_nocut}. 
   \label{fig-flimall_nocut} } 
\end{figure}

\begin{figure}
\plotone{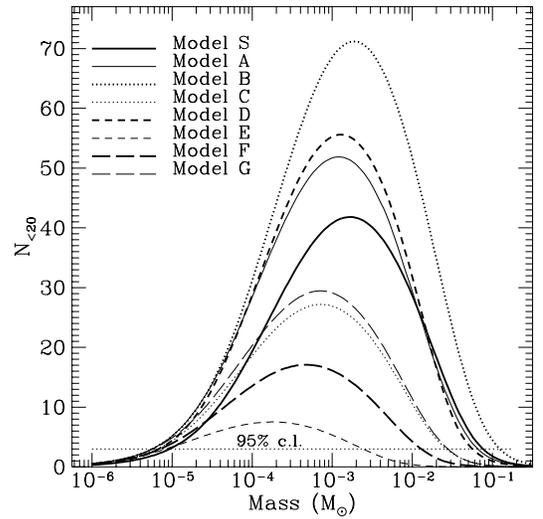}
\caption{ Expected number of events with $\that < 20\,$days 
 for halo models S, A-G, with delta-function Macho mass function. 
 Line coding as Figure~\protect\ref{fig-nexpall_nocut}. 
   \label{fig-nexpall} } 
\end{figure}

\begin{figure}
\plotone{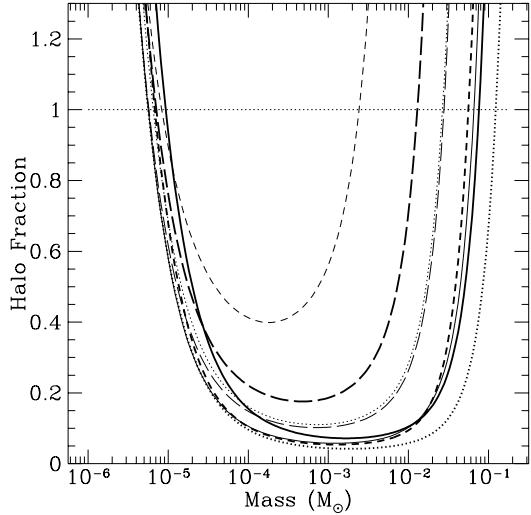}
\caption{ Upper limits (95\% CL) on the fraction of the halo made of 
 Machos for models S, A-G, based on no observed events with $\that < 20\,$days.
 Line coding as Figure~\protect\ref{fig-nexpall_nocut}. 
   \label{fig-flimall} } 
\end{figure}

\begin{figure}
\plotone{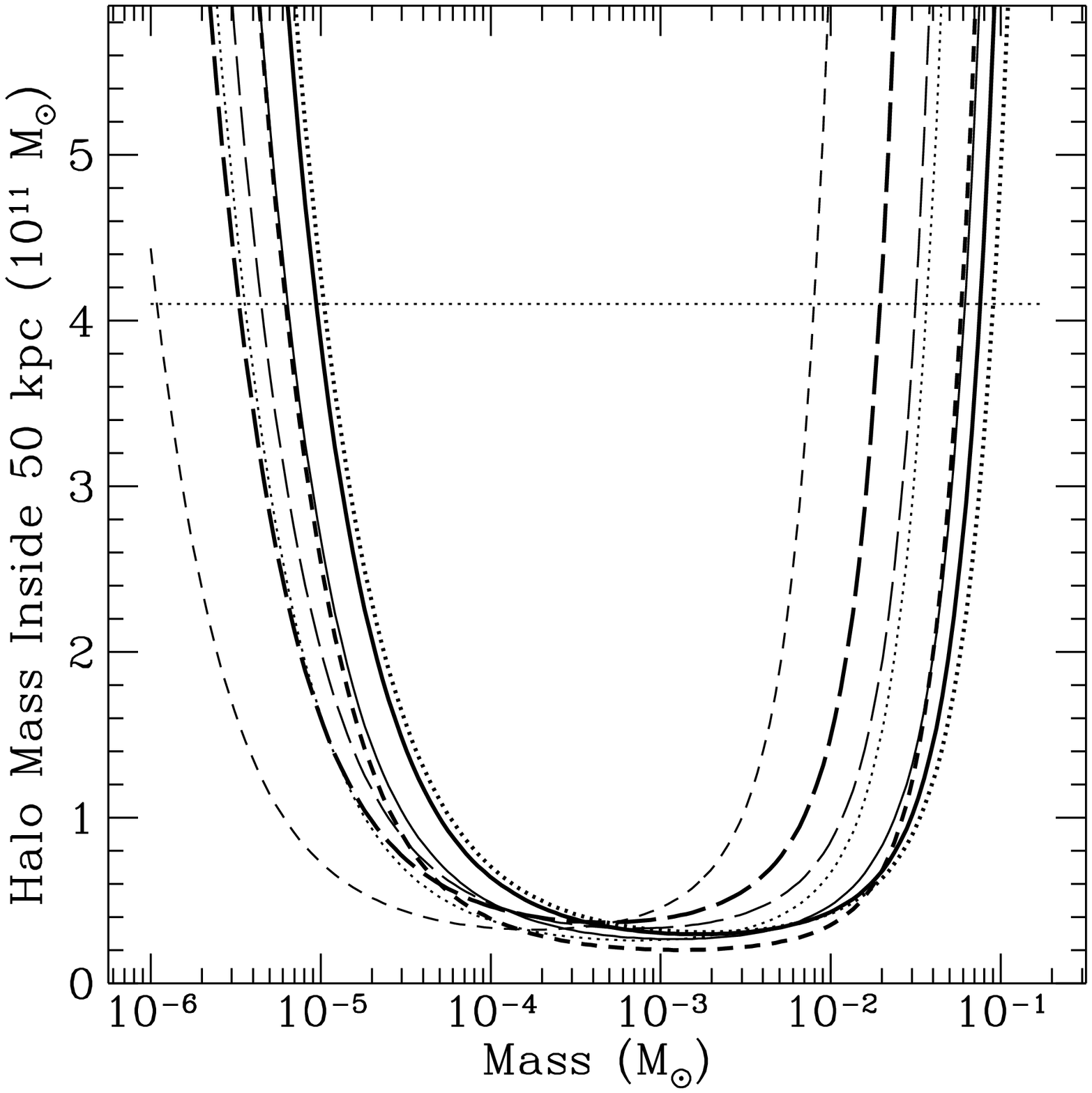}
\caption{ Upper limits (95\% CL) on the the total mass of 
  Machos within $50\,$kpc 
  for models S,A-G,  based on no observed events with $\that < 20\,$days.
 Line coding as Figure~\protect\ref{fig-nexpall_nocut}. 
   \label{fig-mlimall} } 
\end{figure}

The expected number of events $\Nexp$ for each model is
shown in Figure~\ref{fig-nexpall_nocut}, and the resulting
upper limits on halo Macho fraction are shown in
Figure~\ref{fig-flimall_nocut}. 
The expected number of events with $\that < 20$ days $\Nshort$ 
is shown in Figure~\ref{fig-nexpall}; the corresponding
limits on halo fraction $f_{lim} = 3/\Nshort$ are shown in 
 Figure~\ref{fig-flimall}, and the limits on total mass of Machos
interior to $50 \kpc$ are shown in Figure~\ref{fig-mlimall}. 
As in \S~\ref{sec-lowmass}, the latter limit is 
 much stronger for small Macho masses. 
 Except for the unlikely model E, 
 we find that for all the models, Machos between 
 $\sim 2 \ten{-5} \msun$ and $0.01 \msun$ contribute less than 
 $10^{11} \msun$ to the halo mass within $50 \kpc$. 

\begin{figure}
\plotone{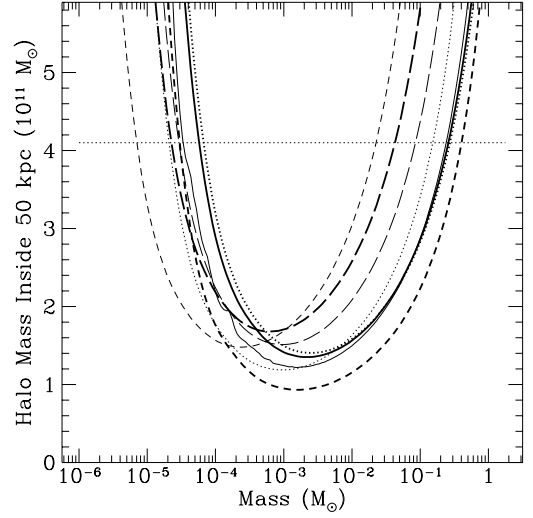}
\caption{ Upper limits (95\% CL) on the 
   total mass of Machos within $50\,$kpc 
  for halo models S, A-G, based upon 8 observed events.
 Line coding as Figure~\protect\ref{fig-nexpall_nocut}. 
   \label{fig-mlimall_nocut} } 
\end{figure}

\begin{figure}
\epsscale{0.8}  
\plotone{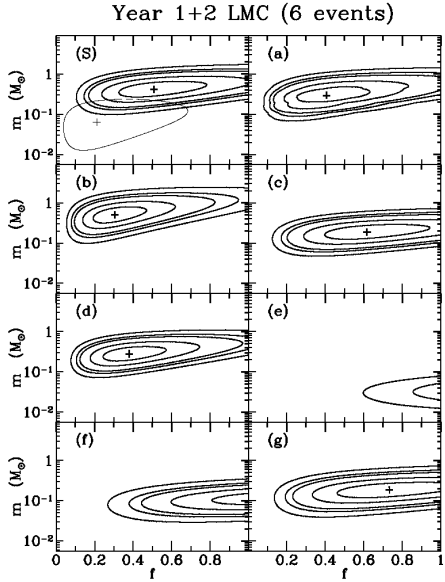}
\caption{ (a) Likelihood contours for Macho mass $m$ and 
 halo fraction $f$ for halo models S, A-G, from the 6-event sample. 
 The $+$ shows the maximum likelihood estimate, and the contours
 enclose regions of $34\%, 68\%, 90\%, 95\%$ and $99\%$ probability. 
 For model S, the 90\% contour for year 1 LMC data (\yrone) is also shown.
 The models are described in \yrone.
   \label{fig-likeall} } 
\end{figure}

The maximum likelihood estimates of Macho
mass $m$, halo fraction $f$, and total mass of Machos
for the 8 models are shown in 
Figure~\ref{fig-likeall} and listed in 
Table~\ref{tab-like}. 
As in \yrone, we find that both the limits $\flim(m)$ and the most
probable halo fraction $f_{ML}$ and Macho mass $m_{ML}$ 
are quite model-dependent, with most probable halo fractions 
ranging from 30\% to 100\%, and most probable
Macho masses ranging from less than $0.1 \msun$ to almost $0.6 \msun$
(in the 6 event sample which is the best to use for this purpose).
Thus the uncertainty introduced due to the halo model is almost as
great as the Poisson errors due to the small number statistics.
However, when we scale the results 
by the total halo mass within $50 \kpc$, giving 
$\Mlim(m) = \flim(m) M_H(50 \kpc)$ and 
$M_{ML}(m) = f_{ML} M_H(50 \kpc)$, 
the results are rather insensitive to 
changes in the halo model.
For limits this is shown in Figure~\ref{fig-mlimall}.
The most probable halo masses are shown in 
column~6 of Table~\ref{tab-like}. 
Thus for the 6 event data set, we find a most probable halo mass in
Machos within $50 \kpc$ of 
$ 2^{+1.2}_{-0.7} \ten{11}\msun$ almost independent
of the halo model.  As discussed in \yrone, this model
independence is not too surprising, 
since our experiment is not sensitive to the total halo mass,
but only to the mass in Machos.  It is worth noting that this is
several times the mass of all known stellar components of the Milky
Way. If the bulk of the lenses are located in the halo, then they
represent the dominant identified 
component of our Galaxy, and a major portion of the dark matter.

\placetable{tab-like} 

\begin{figure}
\epsscale{0.8} 
\plotone{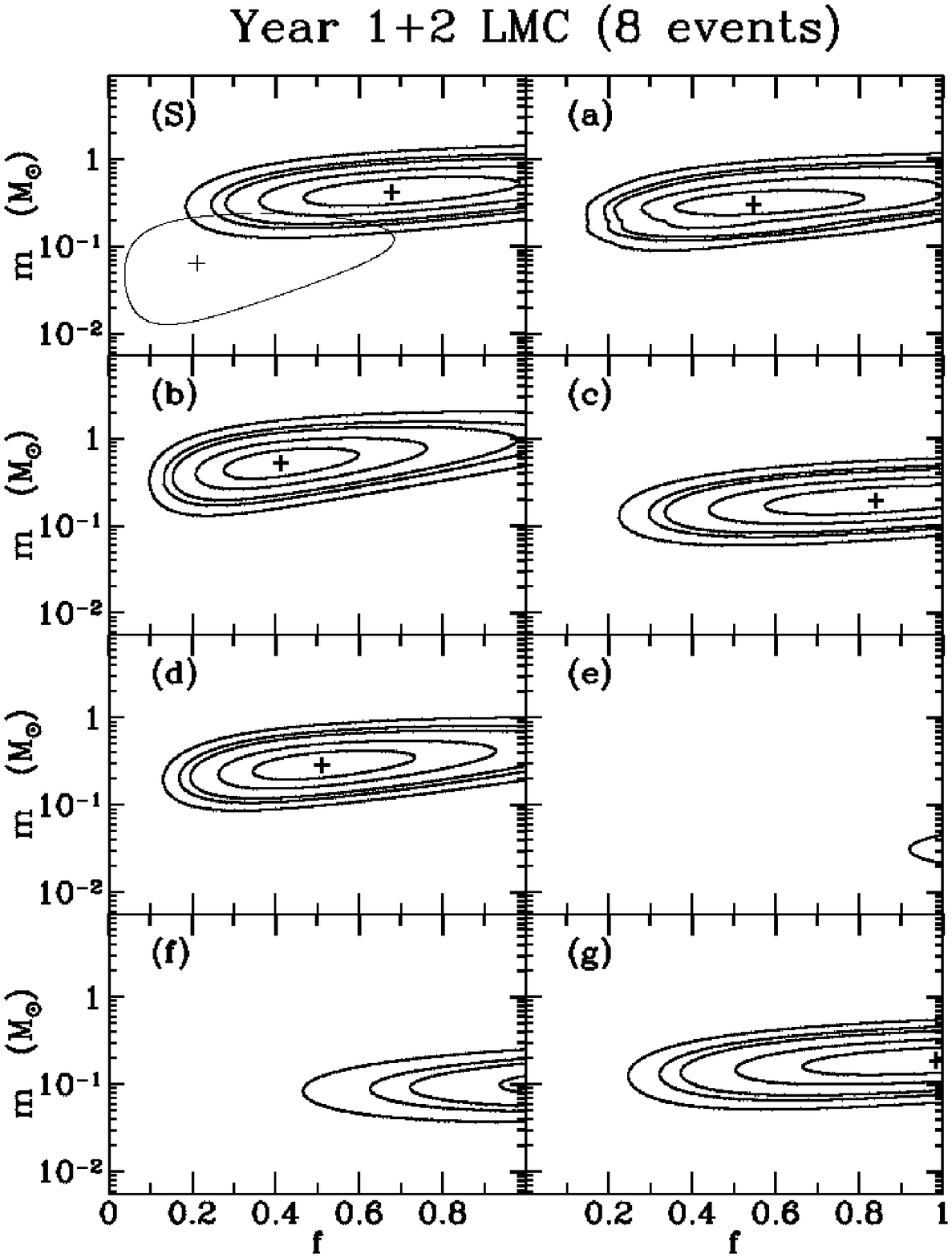}

Figure \protect\ref{fig-likeall}  (b) -- As (a) for the 8-event sample. 
\end{figure}

The model independence of the total halo mass in Machos contrasts
with the model dependence of the most probable lens mass.
One notes that
that the lighter halo models such as C, E, F, G 
have a smaller implied Macho mass $m_{ML}$. 
This arises because lighter halos 
have a smaller velocity dispersion and thus we expect longer
timescale microlensing events for a given mass, i.e. smaller implied
masses for a given observed timescale. 
Also, the lighter halos tend to have larger core radii, 
thus the lenses are on average more distant, 
which also goes in the direction of lengthening 
the events. 
Since $\that \propto \sqrt{m}$, i.e. $m \propto \that^2$, this 
effect can be quite substantial. 
The 95\% CL contours of
models C, E, F, and G overlap considerably the brown dwarf mass
range $m<0.1\msun$.
We note that the implications 
for the formation of the halo could be dramatically different 
in this case.  We also note that rotation of the halo could
lower the expected lens mass range, but these results 
will presented elsewhere.

Also shown in column~7 of Table~\ref{tab-like} are the optical depths
calculated from the halo fraction for each model.  These values and
their confidence intervals are simple to interpret statistically, since
each model provides a distribution of event durations.  Thus the subtleties
discussed in Section~\ref{sec-tau} are absent.  We note that these values
are quite independent of the halo model, and are very close to
the values we obtained in our direct estimates.

\clearpage
\subsection{Other Mass distributions}
\label{sec-imf}

\begin{figure}
\plotone{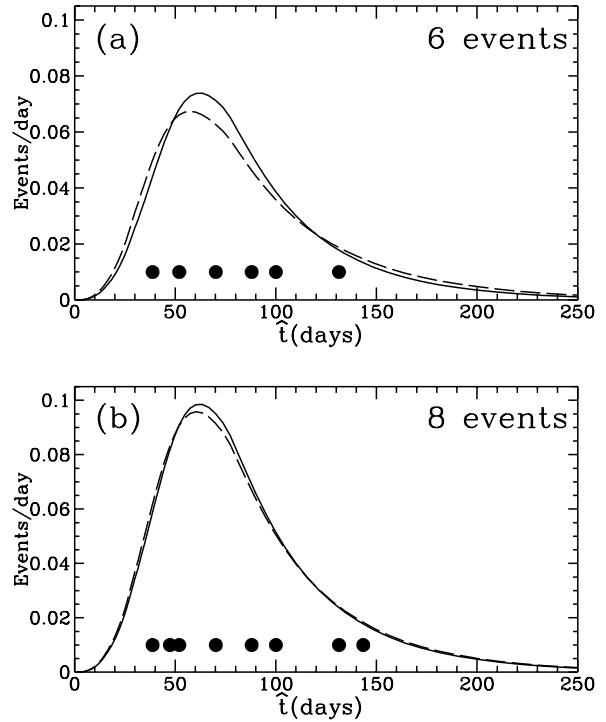}
\caption{ Distribution of durations ($\that$) for maximum likelihood
models.  In part (a) (6 event sample), the solid line is for the best fit
model with a delta-function mass ($m_{2D}=0.41$, $f_{2D}=0.51$), while the
dashed line is for the best power law mass distribution with slope fixed
at $\alpha=-2$ ($m_{min}=0.23$, $f=0.54$).  
The large dots indicate the observed $\that$'s.
The delta-function model has a larger likelihood.
In part (b) (8 event sample), the solid line is for the best fit power
law mass distribution ($\alpha=-3.9$, $m_{min}=0.30$, $f=0.66$), 
which has a slightly
higher likelihood than the dashed line best fit delta-function model
($m_{2D}=0.40$, $f_{2D}=0.65$).
   \label{fig-massdist} } 
\end{figure}

Delta function mass distributions are simple to analyze
and easy to understand, but {\it a priori} are quite unlikely, so
we explored a range of other options.
We repeated our likelihood analysis using a
power-law mass distribution, 
\begin{eqnarray} 
\label{eq-powerm}
  \psi(m) & = A m^\alpha \quad (m_{min} < m < m_{max}) \\
          & = 0          \quad ({\rm otherwise})  \nonumber
\end{eqnarray}
with the normalization constant $A$ determined
from $\int_{m_{min}}^{m_{max}} \psi(m) \, dm  = f$,
where $\psi(m) dm$ is the mass {\it fraction} between $m$ and $m+dm$.
Here we take the slope $\alpha$, cut-off mass $m_{min}$ and halo fraction $f$
as the free parameters in the likelihood function.  
We fixed the maximum mass at $12 \msun$,
since our results did not much depend on its value.
For the 6 event sample we found that the slope became as negative
as possible, and that the minimum mass approached the delta-function mass
model most probable mass.  
Thus the most probable power-law is the one
which most closely approximates the delta-function mass distribution
used earlier!  This can be understood by examining the predicted
distribution of events.   As shown in Figure~\ref{fig-massdist}a,
the most likely delta function mass model ($m_{2D}=0.41$, $f_{2D}=0.51$)
predicts a distribution of durations that matches well the observed 
distribution.  A non-delta mass distribution would necessarily have 
a more spread out duration distribution, which would
be a worse fit to this rather narrow observed event set.  
However, we note that the difference in likelihood between the 
delta-function model and the most likely power-law model with 
$\alpha$ fixed at $-2$  (with its most likely $m_{min}=0.23$ and 
$f= 0.54$) is only 16\%. 
Thus, our data are not really capable
of distinguishing these two cases, as can be seen from the dashed curve
in Figure~\ref{fig-massdist}a.

For the 8 event sample, the most likely distribution 
is a non-delta function distribution with the very steep
slope $\alpha = -3.9$,
$m_{min}=0.30 \msun$, and $f = 0.66$.  This model is favored over
the most likely delta-function model ($m_{2D}=0.40 \msun$,
$f_{2D}=0.65$) by only 2\% in likelihood.   
A single event of longer duration makes the more spread out
duration distribution a slightly better
fit to this wider observed event set, as can be seen in 
Figure~\ref{fig-massdist}b.
The 6 and 8 event sample sets have the same best fit 
delta-function mass because the two samples have approximately the
same mean $\that$.
The two samples do not have the same most likely fraction.  
In fact, by setting the derivative with respect to $f$ 
of Equation~(\ref{eq-like}) equal to zero, we can show that the most likely
$f$ for given $m_{min}$ and $\alpha$ 
is that $f$ which makes the expected number of events 
equal to the observed number of events.
It is for this reason that the plots on Figures~\ref{fig-massdist}a and b
normalize to 6 and 8 expected events respectively.

\subsection{Microlensing by Non-Halo Populations}  
\label{sec-stars}

The implications of the above sections are very striking.
We clearly need to assess 
the significance of the difference between our estimate of the optical depth,
and the optical depth due to known populations of objects.
 In particular, could all of our candidate 
 microlensing events arise from microlensing by 
 known populations of low-mass stars? 
As noted by \citeN{wu94}, \citeN{sahu94}, \citeN{dgmr} and \yrone, 
low-mass stars in the Galaxy and LMC may give rise to microlensing events. 
 These authors find that 
 the optical depth from known stars is
 only $\simlt 10\%$ of that from an all-Macho halo, but is not negligible. 
This may be understood qualitatively 
since the microlensing `tube' is wider in the middle and narrow 
near the Sun and LMC (Griest 1991).

For obtaining confidence intervals, it is better to 
work with microlensing rates rather than optical depth; thus, 
we have repeated the estimates of stellar lensing rates of \yrone, using the 
same model parameters for the thin and thick disks, the spheroid and 
the LMC disk, assuming a Scalo Present Day Mass Function
(PDMF) for all populations, 
and simply updating the efficiency curves and the 
total exposure for our 2-year dataset. 
Two values are given in Table~\ref{tab-stars}
for the LMC disk, first at the center and 
secondly averaged over our fields assuming a 1.6 kpc scale length. 
The values for halo S are also shown for comparison, though a Scalo
PDMF for the halo is implausible.

The results are shown in Table~\ref{tab-stars}; 
the expected number of events 
is $\approx 2 \times$ higher than in \yrone. 
\placetable{tab-stars} 
We see that in our present sample, we would 
expect to observe 0.71 events from stars in the LMC disk, 
0.29 from the Milky Way thin disk, and only 0.075 from the thick disk 
and 0.066 from the  spheroid. This gives a total `background' of 
1.14 expected events from all known stellar populations
and a total background optical depth of $0.54\times 10^{-7}$.

For a Poisson distribution with a mean of 1.14, 
the probabilities to observe 
$\Nobs \ge 3,4,5,6,7$ are $10.7\%, 2.9\%, 0.6\%, 0.11\%, 0.02\%$ 
respectively. 
Thus we see that if only 3 of our events are genuine microlensing, 
the evidence for an excess is modest, whereas if $\ge 4$ of 
our events are microlensing there is very strong evidence for
an excess over stellar lensing alone. 

Since we have 4 `excellent' microlensing candidates, 
it might be tempting to adopt the `null 
hypothesis' that these 4 result from lensing
 by stars (which is only improbable at the $3\%$ level); 
the other candidates are not microlensing (e.g. they are variable stars), 
and then there is only weak evidence for Machos in the halo. 
However, as noted in Section~\ref{sec-events} 
there is a flaw in this argument, in that
the distribution of microlensing magnifications is 
given {\it a priori}, 
and the `excellent' microlensing candidates are preferentially the 
high-magnification ones. 
Thus,  we should expect to find 
 a mixture of `excellent', `good' and `moderate' candidates, and it is 
unlikely that all events would be `excellent' 
(whatever the exact definition). 
 
We can use this argument in two ways: 
firstly, we can add an additional selection cut $\Amax > 2.5$, which 
leaves 4 microlensing candidates (1,4,5,7)  including 3 `excellent' ones; 
this cut reduces the detection efficiency (normalized to 
the event rate with $\Amax > 1.34$) by a factor of 0.71, 
so the expected number of stellar lensing events  
is reduced to $0.81$, and the probability of $\ge 4$ such events is then 
only $0.9\%$. 

Secondly, we can use the results in Table~\ref{tab-umin} 
 for the impact parameter distribution; if we assume that 
 events 1, 4, 5 are microlensing, the observed impact parameter
 distribution is improbable at the 10\% level unless at least 
 one other event is also microlensing. 
 Since the binary event does not count in this table but is 
 almost certainly microlensing, this suggests that we have 
 observed at least  5 microlensing events in total, 
 which is improbable at the $0.6\%$ level from our model
of the stellar populations. 

Thus, we have seen that there appears to be a significant 
excess of lensing events above the expectation from stellar lensing; 
this is true regardless of whether we have observed 4, 5, 6 or 7 real
lensing events, so does not depend sensitively on the lower-quality 
microlensing candidates.  

The main caveat here is that there is some uncertainty 
in the stellar lensing predictions.  For the thick disk and spheroid, 
this is not important since their contributions are very small.
For the thin disk, we have assumed a local column density 
of $50 \msun \pc^{-2}$ for the lens population. 
Dynamical constraints limit the total local 
column density to $\simlt 80 \msun \pc^{-2}$, and 
part of this is in bright stars which will 
not give detectable lensing, and part in gas which does not lens. 
Thus the optical depth from the thin disk 
cannot be much higher than our model value. 

For lensing by LMC stars, the uncertainties are substantially larger. 
The optical depth is proportional to $\Sigma h \sec^2 i$, 
and the event rate is proportional to $\Sigma \sqrt{h} \sec i / \sigma$
where $\Sigma$ is the face-on column density, $h$ is the scale
height and $\sigma$ is the transverse velocity dispersion. 

In Table~\ref{tab-stars} assume $\Sigma_0 = 363 \msun 
 \pc^{-2}$, $h = 250 \pc$, $i = 30^o$ and $\sigma = 25 \kms$
for the LMC, 
 so the important question is how much we could increase these numbers
 without conflicting with dynamical observations. 
For the above $\Sigma_0$ and an exponential scale length of $1.6 \kpc$, 
the LMC rotation curve peaks at $77 \kms$, close
 to the observed value. Thus, it is unlikely that $\Sigma_0$
is much higher than our estimate, but the constraints on 
 $h$ are considerably weaker. 
 However, \citeN{gould-self} has proved a very general relation
  $ \tau = 2\VEV{v^2} \sec^2 i / c^2 $
between the optical depth from a self-gravitating disk and 
the observed line-of-sight velocity dispersion.  For 
 old populations in the LMC, the observed $\VEV{v^2} \simlt 30 \kms$; 
this gives an optical depth of $\simlt 2.6 \ten{-8}$ 
from LMC self-lensing, which is somewhat smaller than our estimate. 

If we assume conservatively that the expected number of events
from LMC stars could be twice the above estimate i.e. 1.42, 
this gives a total of 1.85 expected stellar lensing events 
in our data (this would seem 
to require a large velocity dispersion $\simgt 60 \kms$ for
the old LMC population). 
 The probability of $\Nobs \ge 3,4,5,6,7$ is then
$28\%, 11\%, 4\%, 1\%, 0.3\%$ respectively, so if 5 or more 
events are microlensing there is still a significant excess, 
but if 4 or fewer events are microlensing it is only slightly improbable.  

Thus, it is not impossible to explain our results by stellar lensing
alone, but we have to stretch in several directions simultaneously; we
have to push the observed number down to 4 microlensing events, 
adopt a non-standard LMC
model to get the expected number up close to 2, and bridge the 
remaining gap with  a statistical fluctuation.

\section{Summary and Discussion} 
\label{sec-discuss}

If the observed 
8 candidate events are microlensing, 
the implied optical depth is considerably higher than 
expected from known stars alone. 
In comparison with our \yrone\ results, the implied
event rate is similar ($\sim$ 7 events in 2.1 years 
compared with 3 events in  1.1 years), 
but the longer timescales for the new sample lead 
to considerably higher optical depth and halo mass estimates. 
While the uncertainties are still large, the observed lensing rate is
a significant fraction of that predicted in the standard model of the 
Galactic dark matter halo.   

Using a likelihood analysis to extract information from the 
distribution of event timescales shows a significant model
dependence in the derived halo fraction and the 
individual lens masses.
The total inferred mass in lenses within 50 kpc is quite insensitive to the 
model parameters, however, as is
the optical depth found via likelihood analysis.  These optical depth
estimates and the corresponding confidence intervals are quite close
to our directly estimated optical depth value, and are statistically
simple to interpret.

The experiment's sensitivity to long duration events will
improve over time, and the event tally will presumably
increase as well. Prospects for refining our knowledge of the 
optical depth are therefore promising. 

One natural explanation of the results presented here 
is that a substantial fraction of the 
Galactic dark halo may be made of compact objects. 
We now speculate as to what astrophysical 
objects might be responsible for the observed signal.  
The fact that the observed events have relatively long timescales
suggests that (for standard halo models) 
the lenses have masses above $\sim 0.1 \msun$, with a most probable mass
$\sim 0.5 \msun$. 
If so, they cannot be ordinary 
hydrogen-burning stars since there are strong direct limits 
on such objects from counts of faint red stars 
(e.g. \citeNP{bfgk,hhgc,fgb}); 
thus stellar remnants such as white dwarfs appear to the the most 
obvious possibility.  
However, our exploration of halo models 
also showed that for `minimal' halos, 
the timescales may still be consistent with substellar Machos 
just below $0.1 \msun$. Also, models with a substantial degree of 
halo rotation may lead to smaller mass estimates, 
since the rotation could reduce 
the transverse component of the Macho velocities. Thus, brown 
dwarfs cannot be ruled out as yet, but they would require both a 
non-standard halo model and a mass function concentrated close
to the H-burning limit.  

 There are some theoretical difficulties with the white-dwarf 
 hypothesis (\citeNP{carr94} \& refs. therein); 
 firstly, the initial 
 mass function must be fairly sharply peaked between 
 $\sim 2 - 6 \msun$ to avoid overproducing either 
 low-mass stars (which survive to the present) or 
 high-mass stars (which explode as type-II SNe and overproduce metals). 
 The second difficulty is that the high luminosity of 
 the B and A stars which are the WD precursors 
 may exceed the observed faint galaxy counts 
 \cite{char-silk}, though this may possibly be evaded by dust. 
 
Primordial nucleosynthesis can also provide interesting constraints on
baryonic dark matter. The primordial deuterium abundance is, in principle,
a sensitive indicator of the total baryonic density of the universe. The
observational situation is unclear at present with some measurements indicating
a low deuterium abundance \cite{tytler,burles} and other observations
indicating a high abundance \cite{song,carswell,rugers}. If the low abundance
measurements indicate the actual primordial deuterium abundance, then
a substantial amount of baryonic dark matter must exist. On the other hand,
if the primordial deuterium abundance is low, then there is probably not
enough baryonic dark matter to explain galactic rotation curves. Nevertheless,
as \citeN{rugers} point out, even if the primordial deuterium abundance is
high, primordial nucleosynthesis predicts that the total baryonic mass in
our Galaxy is $\sim 2.4 \times 10^{11} \msun$ 
(for $H_0 = 70\,{\rm km/s Mpc}^{-1}$)
which is consistent with a Macho component of the dark halo as the source
of the bulk of the microlensing events.
 
On the positive side, white dwarfs (along with neutron stars and neutrinos) 
are the only dark matter candidates which are known to exist in 
large numbers. 
Also, it has recently become clear (e.g. \citeNP{white93})
that the mass of hot gas in rich galaxy clusters 
greatly exceeds that in stars, and furthermore this gas is 
 relatively metal-rich with an iron abundance $\sim 0.3$ solar 
\cite{mushotzky}. 
This might suggest that most of the baryons have been processed
through massive stars, which have since died leaving 
a population of remnants and metal-rich gas. This scenario 
has been explored by \citeN{mathews}, who suggest that it may be 
natural to have $40$-$100\%$ of the Galactic dark matter in white
dwarfs. 

The observational limits on the local density of white dwarfs
are a strong function of their age (e.g. \citeNP{liebert}); 
until recently, for ages $\simgt 12$ Gyr 
expected for dark-matter white dwarfs, 
the present limits were an order of magnitude above the halo density. 
These limits were mostly derived from plate-based proper motion surveys, 
and with modern CCD arrays sensitive in the I-band, they could be 
improved by a large factor. 
Very recently, a limit from the Hubble Deep Field 
has been given by \citeN{fgb}; they find that 
white dwarfs with $M_I < 16$ contribute $< 100\%$
of the halo density, and those with $M_I < 15$ contribute $< 33\%$. 

These limits, while tantalizing, are still not stringent enough to
conflict with a large population of old white dwarfs in the halo as an
explanation of our microlensing results; but if such a 
population exists, their local counterparts should be detectable in 
the fairly near future. 

However, it is worth noting that microlensing is sensitive to 
any compact objects, irrespective of composition, as long as they 
are smaller than their Einstein radii $\sim 3$ AU. 
Thus, more exotic objects such as primordial black holes, 
strange stars or `shadow stars' are possible, 
and these would be virtually undetectable by direct searches. 

There are a number of prospects for clarifying the origin of our events: 
firstly, we have now implemented real-time data processing 
for most of our fields, so we hope that most future 
events should be detected in real time. 
Follow-up observations such as frequent high-precision photometry 
 and spectroscopy should help to check that our events are
not due to intrinsic stellar variability. 
For example, our second
 LMC `alert' event 96-LMC-1 was detected on 1996 February~11 and 
announced in IAU Circular~6312; follow-up at CTIO shows a good fit 
to microlensing, with $\Amax = 2.4$ and $\that = 90$ days. 

Secondly, we can discriminate between microlensing by halo or LMC 
objects using the spatial distribution of events, 
as in \S~\ref{sec-dist}. 
Although the current sample does not provide a strong test, 
enlarging the sample by a factor $\sim 2-3$ should clarify this. 
Our project will continue until 1999 December, and
since late 1994 we have modified our observing strategy to 
observe more fields less frequently than the current sample. 
This should increase the detection rate for events with $\that \simgt 
30$ days, and since the additional fields are further from the
center of the LMC, it will also provide a better `lever arm' 
for testing the spatial distribution. 
Also the EROS and OGLE groups plan to have new 
telescopes operating in Chile by late 1996, which should 
increase the detection rate. 

Thirdly, it should be possible to improve the predictions of the 
microlensing rate from LMC stars by direct observation; for example, 
radial velocity measurements for a large sample of LMC RR Lyrae 
stars would provide an important check of the dynamics of 
the old LMC population. If these show a velocity dispersion 
 $\simlt 60 \kms$, it would exclude LMC stars as the main contribution
to the observed lensing \cite{gould-self}. 

Finally, the ideal method for locating the lens population is
to get `parallax' measurements 
by following up the events from a small telescope in Solar orbit 
\cite{gould-2sat,gould-1sat}. 
This measures the velocity of the lens projected to the Solar system, 
which provides a definitive proof of microlensing, 
and can discriminate between disk, halo and 
LMC lenses on an event-by-event basis. 

To summarize, we have two main conclusions: firstly, Machos in the mass range
$\sim 10^{-4}$ to $0.03 \msun$ do not contribute significantly to
the galactic dark matter. 
Secondly, our results 
indicate a microlensing optical depth of $\approx 3 \ten{-7}$
or a Macho mass within 50 kpc of $\approx 2 \ten{11} \msun$. This provides
evidence that Machos with masses in the range $0.05 - 1 \msun$ contribute
a substantial fraction of the galactic dark halo. 
Continued observations from this and other projects should clarify
this in the next few years.

\acknowledgments
\section*{Acknowledgments}

We are very grateful for the skilled support given our project 
by the technical staffs at the Mt.~Stromlo and CTIO Observatories,
and in particular we would like to thank Mr. S. Chan and Mr. G. Thorpe
for their invaluable assistance in obtaining the data.  
We thank the NOAO for making nightly use of the CTIO 0.9~m telescope
possible.
We thank S. Rhie for contributing to the binary fitting
code used for Figure~\ref{fig-bin}.  
Work performed at LLNL is supported by the DOE under contract W7405-ENG-48.
Work performed by the Center for Particle Astrophysics personnel 
is supported in part by the Office of Science and Technology Centers of
NSF under cooperative agreement AST-8809616.
Work performed at MSSSO is supported by the Bilateral Science 
and Technology Program of the Australian Department of Industry, Technology
and Regional Development. 
WJS is supported by a PPARC Advanced Fellowship. 
KG acknowledges support from DOE Outstanding Junior Investigator,
Alfred P. Sloan, and Cottrell awards. 
CWS thanks the Sloan, Packard and Seaver 
Foundations for their generous support. 

\appendix
\section*{Appendix A}

We believe that the blending efficiencies shown in Figure~\ref{fig-eff} are
a reasonably accurate estimate of our
actual microlensing detection efficiencies,
$\eff$, but it is also worthwhile to establish an absolute upper limit on our
detection efficiency. An absolute upper limit on $\eff(\that)$ translates into
a firm lower limit on the microlensing optical depth that will be free of
systematic errors due to blending effects.
 
The upper limit efficiency that we will consider will be an
``optical depth" efficiency, $\eff_\tau(\that)$,
rather than the event detection efficiency
described in the previous section. The optical depth, $\tau$, is defined to
be the probability that a given source star is located within the
Einstein radius of any lensing object, or equivalently the probability,
$P(A \geq 1.34)$, that
a lens has magnified the source star by a factor $\geq 1.34$. One could
also consider the probability that a star is magnified by a factor of
more than some threshold $A_T$. The optical depth would then be given by
$P(A>A_T)/u_T^2$ where $u_T$ is related to $A_T$ by
$A_T = (u_T^2 + 2)/(u_T \sqrt{u_T^2+4})$. For a microlensing experiment
which observes many stars for a finite period of time this becomes
\begin{equation}
\label{eq-tau_obs}
\eqnum{A1} 
\tau_{\rm obs} = {1\over E} \sum_i {T_i(A>A_T)\over u_T^2 \epsilon_i} \ ,
\end{equation}
where $E$ is the total stellar exposure (in star-years) and $T_i(A>A_T)$
is the amount of time that event $i$ is magnified by a factor larger than
$A_T$, and $\epsilon_i$ is the efficiency for detecting
event $i$. Equation~(A1) is correct assuming unblended stellar
images, but now let us consider the effect of blending. In cases when
our lightcurves are
derived from blends of two or more unresolved stellar images, there are
two effects: a) The detection threshold is effectively raised because the
lensed star must be magnified by a larger factor to compensate for the
fact part of the ``source" is not lensed, and b) there is more than one
source that can be lensed. Clearly, these two effects tend to cancel.
We will show, however, that as long as $A_T > 1.34$ ($u_T < 1$),
the effect of blending will always be to {\it reduce} the optical
depth estimate of equation (A1).
 
\begin{figure}
\plotone{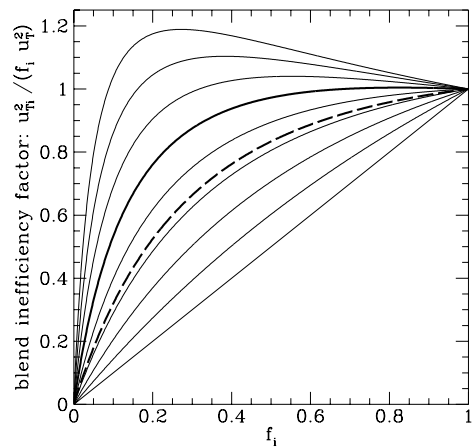}
\caption{ The blend inefficiency factor: $u_{Ti}^2/(f_i u_T^2)$ is plotted
as a function of $f_i$ for $u_T = 0$ (bottom), 0.2, 0.4, 0.6, 0.661
(dashed), 0.8, 1.0 (thick), 1.2, 1.4 and 1.6 (top). 
 The dashed line corresponds to our event detection
threshold of $A_T = 1.75$. These curves indicate how much the optical
depth efficiency is reduced by blending as a function of $f_i$ and $u_T$.
For $u_T<1$ blending always {\it decreases} the optical depth efficiency.
   \label{fig-blend_frac} } 
\end{figure}

Let's consider a blended stellar image consisting of $N$ stars each
contributing a fraction, $f_i$, of the total flux so that $\sum f_i = 1$.
We'll try to estimate the optical depth using equation~(A1),
and we will see how the contribution of the image blend to
equation~(\ref{eq-tau_obs}) compares to the contribution of an
unblended source. Each member of the image blend now has a different
effective magnification threshold given by $A_{Ti}-1 = (A_T-1)/f_i$.
These effective magnification thresholds can be translated into
threshold radii, $u_{Ti}$ using equation~(\ref{eq-amp}). Then, the contribution
of all the stars in the blend to the optical depth is just given by the
ratio of the sum of the areas of the circles of radius $u_{Ti}$ for
each actual star to the area of the single circle corresponding to the
unblended source that was originally assumed. This means that the
apparent microlensing optical depth is modified by the multiplicative factor
$\sum u_{Ti}^2/u_T^2$. This is just the change due to blending of the
area on the sky that can have a (blended) magnification $> A_T$.
Figure \ref{fig-blend_frac} shows the blend inefficiency factor,
$u^2_{Ti}/(f_i u^2_T)$, as a function of $f_i$ and $u_T$. For $u_T < 1$,
this factor is always $< 1$. This means that each blended star will contribute
a fraction $< f_i$ toward the total optical depth as compared to the
optical depth in the unblended case. Since $\sum f_i = 1$, this implies that
the blended stars will contribute less to the optical depth than an unblended
star would. Thus, blending can only serve to decrease the
observed microlensing optical depth.
 
\placefigure{fig-blend_frac}

There are a few caveats regarding applying this argument to our experiment.
First, we do not use equation~(A1) to determine our
measured optical depth. Instead, we use the $\that$ values from unblended
microlensing fits which are shown in Table~\ref{tab-events}
and Figure~\ref{fig-events}. Inspection of
this figure indicates that these fit values do not seriously underestimate
the time that the stars are magnified by more than (say) 1.75. Thus, this
detail should not affect our conclusion.
Another caveat that we must consider is that this argument implicitly assumes
that blending does not make events easier to detect. The one case in which
blending {\it does} make an event easier to detect is the case of events
with $\that > 300$ which are excluded by our timescale cuts. With a blended
event, the fit $\that$ values from the unblended fits used in our event
selection analysis generally underestimate the actual $\that$ values,
so an event with $\that > 300$ days might appear to be an (unblended) event
with $\that < 300$ days which would pass our cuts.
$\eff(\that)$ only enters the optical depth
calculation evaluated at the measured $\that$ values, so
this point has no effect
on our optical depth estimate because our longest detected event
has $\that < 150$ days.

%
%
\clearpage

%







\onecolumn 


\clearpage

\begin{deluxetable}{rccc}  
\tablecaption{ Field Centers \label{tab-fields} \tablenotemark{a} }
\tablewidth{0pt}
\tablehead{
\colhead{Field No.} & 
  \colhead{ Center: RA } & \colhead{Dec (2000)}  & \colhead{Observations} 
}  

\startdata
 1 & 05 05 21.3 & -69 05 13 & 650 \nl
 2 & 05 12 58.2 & -68 28 16 & 456 \nl
 3 & 05 22 36.3 & -68 26 00 & 416 \nl
 5 & 05 11 13.0 & -69 41 00 & 501 \nl
 6 & 05 19 54.4 & -70 18 49 & 509 \nl 
 7 & 05 28 47.4 & -70 29 46 & 618 \nl 
 9 & 05 11 07.4 & -70 23 42 & 446 \nl
10 & 05 04 20.8 & -69 53 56 & 363 \nl
11 & 05 37 07.6 & -70 32 34 & 567 \nl
12 & 05 45 42.6 & -70 34 44 & 435 \nl
13 & 05 19 41.7 & -70 52 38 & 383 \nl
14 & 05 35 58.1 & -71 09 20 & 379 \nl
15 & 05 45 46.4 & -71 15 31 & 398 \nl 
18 & 04 57 49.1 & -68 55 46 & 313 \nl
19 & 05 06 10.3 & -68 18 45 & 378 \nl 
47 & 04 52 56.6 & -68 00 27 & 300 \nl
77 & 05 27 27.1 & -69 45 41 & 785 \nl
78 & 05 19 17.6 & -69 43 39 & 745 \nl
79 & 05 12 53.1 & -69 05 15 & 684 \nl
80 & 05 22 55.0 & -69 04 00 & 634 \nl
81 & 05 36 19.0 & -69 48 00 & 441 \nl
82 & 05 33 06.0 & -69 03 00 & 426 \nl
\enddata
\tablenotetext{a} { This table lists the 22 well-sampled 
 fields used in the current
 analysis. We observe 82 LMC fields in total, but the remaining 60 
 were observed much less often ($\sim 60$ observations each) 
 in the first 2 years. 
 } 
\end{deluxetable}

\clearpage
\begin{deluxetable}{cccc}  
\tablecaption{ Selection Criteria \label{tab-cuts} }
\tablewidth{0pt}
\tablehead{
 \colhead{Description} & 
  \colhead{Old} & \colhead{New}  & \colhead{Change} 
}  

\startdata
 Crowding   & $f_{CRD} < 1.67 $ & 
 $f_{CRD} < \left(\Delta\chi^2 /(\chi^2/\ndf)\right)^{10/9}/520$ & Loosened \nl
 SN echo   & $10'\times10'$ square excl. & 
       $10'\times10'$ square excl.  & Unchanged \nl
 Coverage & $\geq 3$ pts.~on rise \& fall & $\geq 1$ pt.~on rise \& fall &
        Loosened \nl
 High points & 8 points $> 1\sigma$ high & 
        6 points $>1\sigma$ high & Loosened \nl
 Baseline fit & $\chi^2_{ml}/\ndf < 3$ & $\chi^2_{ml-out}/\ndf < 4$ & 
        Loosened \nl 
 Peak fit   & $\chi^2_{peak}/\ndf < 4$ & 
          $\Delta\chi^2 / (\chi^2_{peak}/\ndf) > 200$  &  Loosened \nl
 Chromaticity & $P_{\rm achrom} < 0.997$  & None  & Loosened \nl 
 Peak significance  & $\Delta\chi^2 /(\chi^2_{ml}/\ndf) > 200$   & 
          $\Delta\chi^2 /(\chi^2_{ml}/\ndf) >500$ & Tightened \nl
 Amplification & $\Amax > {\rm max}(1.50, 1+2\overline{\sigma}) $ & 
         $\Amax > {\rm max}(1.75, 1+2\overline{\sigma}) $ & Tightened \nl
\enddata
\end{deluxetable} 

\clearpage
\begin{deluxetable}{rcccccccccc}  
\tablecaption{ Candidate Microlensing Events \label{tab-events} }
\tablewidth{0pt}
\tablehead{
\colhead{Event \tablenotemark{a}} & 
  \colhead{RA} & \colhead{Dec (2000)}  & \colhead{V} & 
   \colhead{V-R}  & \colhead{$\tmax$} & \colhead{$\that$} & 
    \colhead{$\Amax$}    & \colhead{$f_{0R}$}  & \colhead{$f_{0B}$}  & 
    \colhead{$\chi^2/\ndf$}
}  

\startdata
1a & 05 14 44.3  & -68 48 01 &   19.6 & 0.6  & 57.08(3)  & 34.7(3)  &
    7.2(1) & 93.4(3) & 63.0(3) & 1.420  \nl
1b & 05 14 44.3  & -68 48 01 &   19.6 & 0.6  & 57.26(4)  & 34.3(3)  &
    7.5(3) & 77.7(2) & 47.8(2) & 1.134  \nl
4 & 05 17 14.6 & -70 46 59 &  20.0 & 0.2  & 647.2(2) & 46(2) & 
    3.00(4)  & 35.6(2) & 40.6(3) & 1.416  \nl
5 & 05 16 41.1  & -70 29 18 &  20.7 & 0.4  & 24.0(3) & 82(2) & 
   58(5) & 26.5(4) & 20.5(3) & 1.680  \nl
6 & 05 26 14.0 & -70 21 15 &  19.6 & 0.3  & 197.5(7) & 87(4) & 
   2.14(4)  & 57.1(5) & 59.3(5) & 0.873 \nl
7 & 05 04 03.4 & -69 33 19 &  20.7 & 0.4  & 463.0(3) & 115(3) & 
    6.16(10)  & 23.2(3) & 23.3(3) & 1.447  \nl
8 & 05 25 09.4 & -69 47 54 &  20.1 & 0.3  & 388.4(5) & 62(2) & 
    2.24(5)  & 35.1(4) & 39.4(3) & 2.218  \nl
9\tablenotemark{b} & 05 20 20.3 & -69 15 12 &  19.3 & 0.3  & 
  597.1(8) & 147(3) &  1.86(15)  & 77.5(4) & 81.0(3) & 6.868  \nl
10 & 05 01 16.0 & -69 07 33 &  19.4 & 0.2  & 205.3(3)  & 42(1) & 
    2.36(5)  & 67.4(3) & 79.8(3) & 1.982  \nl
11 & 05 34 21.8 & -70 41 07 &  21.5  & 0.4   & -8.6(3)  & 266(9) & 
    11.9(4)  & 74(2) & 64(2) & 2.964  \nl
12a & 05 33 51.7 & -70 50 59 &  21.2  & 0.3  & -10.0(3)  & 138(5) & 
    7.2(4)  & 14.3(3) & 14.6(3) & 1.487  \nl
12b & 05 33 51.7 & -70 50 59 &  21.2  & 0.3  & -11.4(8)  & 162(8) & 
    6.8(2)  & 12.6(3) & 11.5(2) & 1.536  \nl
\enddata
\tablenotetext{a} {Events 1-3 appeared in \yrone, but 2 \& 3 are not in
  the sample used here. 
 We number the current sample 1, 4 \ldots 12 to avoid ambiguity. } 
\tablenotetext{b} {Event 9 is the binary microlensing event; the parameters
 here are those resulting from a single-lens fit, and are not 
  strictly appropriate.} 
\tablenotetext{} {The magnitudes and colors are based on an 
 approximate transformation from our non-standard passbands. 
 Time of peak magnification $\tmax$ is in JD - 2,449,000. 
 The microlensing fit parameters $\tmax$,$\that$,
 $\Amax$ are defined in eq.~\protect\ref{eq-mlfit}. 
 Units of $f_{0B}, f_{0R}$ are arbitrary but are given for comparison
 with Table~\protect\ref{tab-blend}. 
 The figure in brackets gives the 
  formal $1\sigma$ uncertainty of the least significant digit(s).}
\end{deluxetable} 

\clearpage
\begin{deluxetable}{rcccccccc}  
\tablecaption{ Microlensing Fits with Blending \label{tab-blend} }
\tablewidth{0pt}
\tablehead{
\colhead{Event} & 
   \colhead{$\tmax$} & \colhead{$\that$} & 
    \colhead{$\Amax$}    & \colhead{$f_{lR}$}  & \colhead{$f_{lB}$}  & 
    \colhead{$f_{uR}$}  & \colhead{$f_{uB}$}  & 
    \colhead{$\chi^2/\ndf$}
}  

\startdata
1a &  57.08(3)  & 34.9(4)  &
    7.25(16) & 93(2) & 62.0(9) & 0(2) & 1.2(1.0) & 1.422  \nl
1b &  57.26(4)  & 34.3(8)  &
    7.5(3) & 78(3) & 48(2) & 0(3) & 0(2) & 1.136  \nl
4 & 646.8(2) & 79(13) & 
    6.4(1.2)  & 14(4) & 15(4) & 21(4) & 26(4) & 1.383  \nl
5 &  24.0(3) & 111(3) & 
   220(70) & 15.4(5) & 15.6(4) & 17.4(7) & 0.0(2) & 0.965  \nl
6 & 197.5(7) & 94(3) & 
   2.37(9)  & 47(3) & 51(2) & 10(3) & 8(2) & 0.875 \nl
7 & 463.0(3) & 128(3) & 
    7.1(2)  & 18.1(5) & 21.7(3) & 6.6(7) & 0.0(2) & 1.318  \nl
8 & 388.4(5) & 62(3) & 
    2.28(11)  & 35(3) & 38(2) & 0(3) & 2(2) & 2.227  \nl
9\tablenotemark{a} & 603.04(2) & 143.4(2) &  
    ()  & 20.57(13) & 14.48(8) & 58.8(3) & 68.82(15) & 1.755  \nl
10 & 205.3(3)  & 42.7(1.3) & 
    2.40(5)  & 66(2) & 78(3) & 2(2) & 2(3) & 1.989  \nl
\enddata
\tablenotetext{a} {For the binary microlensing event (9) the fit
 parameters given are for the binary lens fit described in 
 \cite{macho-bennett96,macho-lmcbinary}
Not all of the these parameters are appropriate for this fit.}
\tablenotetext{} {
 Time of peak magnification $\tmax$ is in JD - 2,449,000. 
 The microlensing fit parameters $\tmax$, $\that$, and
 $\Amax$ are defined in eq.~\protect\ref{eq-mlfit}. 
 The figure in brackets gives the 
  formal $1\sigma$ uncertainty of the least significant digit(s).}
\end{deluxetable} 

\clearpage
\begin{deluxetable}{ccc}  
\tablecaption{ Impact Parameter Distribution \label{tab-umin} }
\tablewidth{0pt}
\tablehead{
\colhead{Event Subset} & 
  \colhead{$\VEV{\umin}$ } & \colhead{$P(<\umin)$ \tablenotemark{a} }
}  

\startdata
1,5       & 0.080 & 0.032 \nl 
1,5,7     & 0.109 & 0.024 \nl 
1,4,5,7   & 0.169 & 0.061 \nl 
1,4-7     & 0.237 & 0.194 \nl 
1,4-8     & 0.279 & 0.343 \nl 
1,4-8,10  & 0.304 & 0.466 \nl 
1,4,5     & 0.170 & 0.097 \nl 
\enddata
\tablenotetext{a} {This is the probability that, given a number of
 detected events equal to that in the subsample, the 
 mean $\umin$ would be smaller than the observed value in the previous 
 column.} 
\end{deluxetable} 

\clearpage
\begin{deluxetable}{cccccc}  
\tablecaption{ Spatial Distribution Statistical Tests \label{tab-space} }
\tablewidth{0pt}
\tablehead{
\colhead{Events} & \colhead{bar bias } & \colhead{KS distance} &
\colhead{$P_{KS}$} & \colhead{$P_{KS}$ (1-sided)} &\colhead{$P_{Wilcoxon}$}
}  

\startdata
8 &  1 & -0.268 & 0.546 & 0.273 & 0.83 \nl 
8 &  4 & +0.233 & 0.717 & 0.358 & 0.35 \nl 
8 & 12 & +0.386 & 0.140 & 0.070 & 0.12 \nl 
6 &  1 & -0.258 & 0.754 & 0.388 & 0.68 \nl 
6 &  4 & +0.318 & 0.495 & 0.243 & 0.28 \nl 
6 & 12 & +0.417 & 0.176 & 0.088 & 0.12 \nl 
\enddata
\tablenotetext{} {The results of
the Kolmorogov-Smirnov 2-sided and 1-sided tests, and that of
the Wilcoxon test on the distribution of distances from the
lensing event locations to the center of the LMC bar for our 8 and 6
event samples are shown for models with `bar biases' of
1 (halo lensing), 4, and 12. The `bar bias' refers to the factor by
which the microlensing optical depth in the bar exceeds the optical
depth outside the bar. 
The sign on the KS distance is positive if the
observed distribution is more uniform than the model.
The 1-sided KS probabilities refer to 1-sided KS tests of
bias$\,=1$ models against the bias$\,>1$ hypothesis
and the bias$\,=4$ or 12 models against the bias$\,<4$ or bias$\,<12$
hypotheses.
The Wilcoxon probability is that the bar distances
from the observations are {\it not}
systematically greater than those predicted
by the models.}
\end{deluxetable}

\clearpage
\begin{deluxetable}{rccc}  
\tablecaption{ Single Event Optical Depths\label{tab-blthat} }
\tablewidth{0pt}
\tablehead{
\colhead{\qquad Event \qquad} & 
   \colhead{\qquad $\that$ \qquad } &
   \colhead{\qquad $\that_{bl}$ \qquad } &
   \colhead{\qquad $\tau_1$ \qquad } 
}  

\startdata
1 & 34.7  & 38.8 & $1.8 \ten{-8}$ \nl
4 & 46    & 52   & $2.3 \ten{-8}$ \nl
5 & 82    & 88   & $3.5 \ten{-8}$ \nl 
6 & 87   & 100   & $4.1 \ten{-8}$ \nl
7 & 115  & 131   & $6.0 \ten{-8}$ \nl
8 & 62   &  70   & $2.8 \ten{-8}$ \nl
9 & 147  & 143   & $6.6 \ten{-8}$ \nl
10 & 42  &  47   & $2.1 \ten{-8}$ \nl
\enddata
\tablenotetext{} {The quantity $\that_{bl}$ is the average actual event 
timescale for
events in our Monte Carlo calculations which are detected with an
unblended fit timescale of $\that$. $\that_{bl}$ can be used as an
unbiased estimator of the actual $\that$ value. For the binary event, \# 9,
the blended fit $\that$ value is used instead. $\tau_1$ indicates the
contribution of each event to the total microlensing optical depth,
computed using equation~(\ref{eq-tau1}).}
\end{deluxetable}

\clearpage
\begin{deluxetable}{ccccccccc} 
\tablecaption{ Optical Depth Confidence Intervals \label{tab-tau} }
\tablewidth{0pt}
\tablehead{
\colhead{$\eff$ type} & \colhead{\# of events} & 
  \multicolumn{7}{c}{$\tau (10^{-7})$ for confidence level:} \nl
\multicolumn{2}{c}{} & 
                       \colhead{0.025} & \colhead{0.05} & \colhead{0.16} & 
 \colhead{measured} & \colhead{0.84} & \colhead{0.95} & \colhead{0.975}
}  

\startdata

{\bf photometric }     & {\bf 8} & {\bf 1.24} & {\bf 1.47} & {\bf 1.99}
  & {\bf 2.93} & {\bf 4.30} & {\bf 5.28} & {\bf 5.82}\nl
sampling      & 8 & 1.12 & 1.31 & 1.77 & 2.59 & 3.79 & 4.66 & 5.12\nl
photometric $\that$-fit & 8 & 1.33 & 1.55 & 2.10 & 3.06 & 4.49 & 5.53 & 6.07\nl
sampling $\that$-fit &8&1.18& 1.38 & 1.85 & 2.71 & 3.95 & 4.85 & 5.32\nl
sampling no $\that$-cor. &8 & 1.00 & 1.17 & 1.60 & 2.34 & 3.44 & 4.23 & 4.67\nl
photometric      & 6 & 0.77 & 0.93 & 1.33 & 2.06 & 3.18 & 4.00 & 4.44\nl
sampling no $\that$-cor. &6 & 0.61 & 0.74 & 1.04 & 1.59 & 2.43 & 3.06 & 3.40\nl
\enddata
\tablenotetext{} { The table entries 
 show limits at various confidence levels on 
 the microlensing optical depth $\tau$ in units of $10^{-7}$, 
with different assumptions for the detection efficiency $\eff$, 
different $\that$ corrections (due to blending),
and 6 or 8 events assumed to be microlensing. Our photometric efficiencies 
are considered the most accurate estimate of the true microlensing
detection efficiency, and the average $\that$ correction
is preferred because it is unbiased. The photometric efficiency
results with the average $\that$ correction are displayed in bold face type.
The `no $\that$ correction' values are to be considered extreme lower limit
values. }
\end{deluxetable}

%
%
\clearpage

\begin{deluxetable}{cccccccccc}  
\tablecaption{ Maximum Likelihood Fits \label{tab-like} }
\tablewidth{0pt}
\tablehead{
\colhead{Events \tablenotemark{a} } & 
\colhead{Model \tablenotemark{b} } & 
\colhead{Halo} &
  \colhead{$ m_{ML} (\msun)$ \tablenotemark{c} } & 
   \colhead{$f_{ML} $} & 
   \colhead{$ f_{ML} M_{H} (10^{10}\msun)$ }  & 
   \colhead{$\tau_{ML} (10^{-8}) $} 
}  

\startdata
6 & S & med. & $0.46^{+0.30}_{-0.17}$  & $0.50^{+0.30}_{-0.20}$  
& $20^{+12}_{-7} $ & $24^{+14}_{-8} $ \nl
\nl
6 & A & med.  &$0.32^{+0.25}_{-0.11} $ & $0.41^{+0.25}_{-0.17}$  
& $19^{+11}_{-8} $ & $23^{+14}_{-9} $ \nl
\nl
6 & B & large & $0.55^{+0.38}_{-0.21} $ & $0.30^{+0.19}_{-0.12}$  
& $22^{+14}_{-9} $ & $24^{+15}_{-10} $ \nl
\nl
6 & C & small & $0.21^{+0.12}_{-0.08} $ & $0.61^{+0.36}_{-0.25}$  
& $14^{+8}_{-6} $ & $18^{+11}_{-8} $ \nl
\nl
6 & D & E6 & $0.31^{+0.18}_{-0.11} $ & $0.37^{+0.22}_{-0.15}$  
& $17^{+10}_{-7} $ & $22^{+13}_{-8} $ \nl
\nl
6 & E & max disk & $0.04^{+0.02}_{-0.01} $ & $2.8^{+?}_{-?}$  
& $22^{+?}_{-?} $ & $24^{+?}_{-?} $ \nl
\nl
6 & F & big disk & $0.13^{+0.08}_{-0.05} $ & $1.2^{+0.70}_{-0.47}$  
& $25^{+14}_{-10} $ & $23^{+13}_{-9} $ \nl
\nl
6 & G & big disk & $0.21^{+0.12}_{-0.08} $ & $0.71^{+0.44}_{-0.29}$  
& $23^{+14}_{-9} $ & $23^{+14}_{-10} $ \nl
\noalign{\hrule} \nl
8 & S & med. & $0.45^{+0.24}_{-0.15}$  & $0.68^{+0.33}_{-0.23}$  
& $28^{+13}_{-9} $ & $32^{+16}_{-11} $ \nl
\nl
8 & A & med.  &$0.32^{+0.20}_{-0.09} $ & $0.55^{+0.28}_{-0.20}$  
& $25^{+13}_{-9} $ & $31^{+16}_{-11} $ \nl
\nl
8 & B & large & $0.56^{+0.31}_{-0.20} $ & $0.41^{+0.21}_{-0.15}$  
& $29^{+15}_{-11} $ & $33^{+17}_{-12} $ \nl
\nl
8 & C & small & $0.21^{+0.11}_{-0.07} $ & $0.83^{+0.40}_{-0.29}$  
& $19^{+9}_{-7} $ & $25^{+12}_{-9} $ \nl
\nl
8 & D & E6 & $0.31^{+0.16}_{-0.11} $ & $0.50^{+0.25}_{-0.18}$  
& $23^{+12}_{-8} $ & $30^{+15}_{-11} $ \nl
\nl
8 & E & max disk & $0.04^{+0.02}_{-0.01} $ & $>1$  
& ? & ? \nl
\nl
8 & F & big disk & $0.13^{+0.06}_{-0.04} $ & $1.67^{+0.70}_{-0.52}$  
& $34^{+14}_{-11} $ & $32^{+14}_{-10} $ \nl
\nl
8 & G & big disk & $0.20^{+0.11}_{-0.07} $ & $0.97^{+0.49}_{-0.34}$  
& $31^{+16}_{-11} $ & $32^{+16}_{-11} $ \nl
\nl

\enddata
\tablenotetext{a} { This column shows the number of 
 events assumed to result from halo microlensing. 
 The 8-event sample is 1,4\ldots10 ; the 6-event sample excludes
 events 9 \& 10. }
\tablenotetext{b} {The models are defined as in \yrone; in summary, 
 model S is given by eq.~\ref{eq-stdhalo}, the others are Evans models. 
Model~A is similar to model~S. 
Models B and~C have more and less massive
halos; model~D is similar to~A but flattened to E6. 
Model~E is an extreme maximal disk/minimal halo model, while F and~G 
 are more realistic heavy disk/ light halo models. }
\tablenotetext{c} {Columns 4 \& 5 show the maximum likelihood Macho mass
 and halo fraction from Section~\ref{sec-like}.  
 Columns 6 \& 7 show the implied total mass of Machos 
 within $50 \kpc$ of the Galactic center, 
 and the resulting optical depth.
For model (E) some entries are marked ``?" indicating
that the halo fraction became unreasonably large and 
the numerically calculated error estimates were therefore inaccurate. }
\end{deluxetable}

\clearpage
\begin{deluxetable}{cccccc}  
\tablecaption{ Microlensing by Stars \label{tab-stars} 
    \tablenotemark{a} }
\tablewidth{0pt}
\tablehead{
\colhead{Population } & 
  \colhead{$\tau (10^{-7}) $} & \colhead{$ \VEV{\that}$ (days) } & 
  \colhead{$ \VEV{l} (\kpc) $}  & 
  \colhead{$ \Gamma (10^{-7} {\rm yr^{-1}}) $}  &
  \colhead{$ \Nexp $} 
}  

\startdata
Thin disk   &  0.15  & 112  & 0.96  & 0.62  & 0.29  \nl 
Thick disk  & 0.036  & 105  & 3.0   & 0.16  & 0.075 \nl
Spheroid    & 0.029  & 95   & 8.2   & 0.14  & 0.066 \nl
LMC center  & 0.53   & 93   & 49.8  & 2.66  & (1.19) \nl
LMC average & 0.32   & 93   & 49.8  & 1.60  & 0.71  \nl
Halo S      & 4.7    & 89   & 14.4  & 24.3  & 11.2  \nl
\enddata
\tablenotetext{a} {This table shows microlensing quantities for
 model stellar populations, with Scalo PDMF and the density and 
 velocity distributions given in \yrone. $\VEV{l}$ is the mean 
 lens distance. $\Gamma$ is the total theoretical microlensing
 rate. The expected number of events $\Nexp$ includes our detection
 efficiency averaged over the $\that$ distribution. 
 For the LMC, two rows are shown; firstly at the center, and 
 secondly averaged over the location of our fields; only the
 averaged $\Nexp$ is relevant.  } 
\end{deluxetable} 

\clearpage




\end{document}